\documentclass[aps,reprint]{revtex4-1}

\usepackage{commath}
\usepackage{amsmath}
\usepackage{graphicx}
\usepackage{hyperref}
\usepackage{txfonts}

\draft % marks overfull lines with a black rule on the right

\begin{document}

% Use the \preprint command to place your local institutional report number 
% on the title page in preprint mode.
% Multiple \preprint commands are allowed.
%\preprint{}

\title{Theory of Non-Retarded Ballistic Surface Plasma Waves in Metal Films}
% repeat the \author .. \affiliation  etc. as needed
% \email, \thanks, \homepage, \altaffiliation all apply to the current author.
% Explanatory text should go in the []'s, 
% actual e-mail address or url should go in the {}'s for \email and \homepage.
% Please use the appropriate macro for the type of information

% \affiliation command applies to all authors since the last \affiliation command. 
% The \affiliation command should follow the other information.

\author{Hai-Yao Deng}
\email{h.deng@exeter.ac.uk}
%\author{Eros Mariani$^{1}$}
%\author{Chi-Hang Lam$^2$}
%\email{c.h.lam@polyu.edu.hk}
%\author{Katsunori Wakabayashi$^{3,4}$}
%\email{waka@kwansei.ac.jp}
%\homepage[]{Your web page}
%%\thanks{}
%\altaffiliation{}
\affiliation{Department of Physics and Astronomy, University of Exeter, EX4 4QL Exeter, United Kingdom}
%\affiliation{$^2$Department of Applied Physics, The Hong Kong
%Polytechnic University, Hung Hum, Hong Kong}  
%\affiliation{$^3$Department of Nanotechnology for Sustainable Energy,
%School of Science and Technology, Kwansei Gakuin University, Gakuen 2-1,
%Sanda 669-1337, Japan}
%\affiliation{$^4$National Institute for Materials Science (NIMS), Namiki 1-1, Tsukuba 305-0044, Japan}
% Collaboration name, if desired (requires use of superscriptaddress option in \documentclass). 
% \noaffiliation is required (may also be used with the \author command).
%\collaboration{}
%\noaffiliation

\begin{abstract} 
We present a theory of surface plasma waves in metal films with arbitrary electronic collision rate $\tau$. Both \textit{tangential} and \textit{normal} modes are investigated. A universal self-amplification channel for these waves is established as a result of the unique interplay between ballistic electronic motions and boundary effects. The channel is shown to be protected by a general principle and its properties independent of $\tau$. The effects of film thickness and surface roughness are also calculated. Experimental implications, such as Ferrel radiation, are discussed. 
\end{abstract}

%\pacs{51.10.+y, 52.25.Dg, 52.27.Aj, 73.20.Mf, 73.22.Lp}% insert suggested PACS numbers in braces on next line

\maketitle %\maketitle must follow title, authors, abstract and \pacs

\section{Introduction}
Surface plasma waves (SPWs)~\cite{ritchie1957,ferrell1958,raether1988} are fascinating to a wide spectrum of scientists not only for their fundamental physical properties~\cite{pitarke2007, feibelman1982, pendry1975} but also their promising potential~\cite{int} in a myriad of applications, including microscopy~\cite{benno1988}, sensing~\cite{hoa2007} and nano-optics~\cite{zayats2005,maier2007,ozbay2006,mark2007,barnes2003,ebbesen2008} as well as information processing~\cite{tame2013}. Being charge density waves highly localized about the interface between a metal and a dielectric, SPWs strongly interact and form a bound entity with light that might render an atomic resolution of molecular dynamics~\cite{nie1997}. Nowadays SPWs are pivotal in nano-optics. 

The standard theory of SPWs was delivered shortly after the pioneering work~\cite{ritchie1957} by Ritchie in 1957 and has since been comprehensively discoursed in many textbooks and review articles~\cite{raether1988,pitarke2007,maier2007,sarid2010}. In this theory, the electrical properties of a metal are prescribed with a dielectric function $\epsilon$. To analytically treat $\epsilon$, the simple Drude model or the slightly more involved hydrodynamic model is often invoked~\cite{pitarke2007,harris1971,fetter1986,pendry2013,pendry2014,pendry2015,schnitzer2016}. For either model to be valid, electronic collisions in the metal must be sufficiently frequent so that the electronic mean free path, $l_0=v_F\tau$, where $v_F$ is the Fermi velocity and $\tau$ the thermal charge relaxation time, is much shorter than the SPW wavelength or the typical length of the system~\cite{ziman1960,pines,abrikosov}. The general case with arbitrary $\tau$, especially the collision-less limit, where $\tau\rightarrow\infty$, defies these models and has yet to be entertained. Other models based on \textit{ab initio} quantum mechanical computations~\cite{pitarke2007} are helpful in understanding the complexity of real materials but falls short in providing an intuitive and systematic picture of SPWs underpinned by electrons experiencing less frequent collisions.

The purpose of this paper is to furnish a comprehensive theory for SPWs of ballistically moving electrons. Ballistic SPWs are not only interesting in themselves but could have ramified applications in plasmonics and other arenas. Recently~\cite{deng2017a,deng2017b}, we considered ballistic SPWs in semi-infinite metals. We showed that such waves are intrinsically unstable and possess a universal self-amplification channel that exists irrespective of the value of $\tau$. This result was initially established by examining the charge dynamics~\cite{deng2017a} in the system and later corroborated by an energy conversion analysis~\cite{deng2017b} in the waves. In the present work, we study ballistic SPWs in metal films, which possess two surfaces and are experimentally more realistic and interesting.

In the next section, we specify the system under consideration and state our main results. Some preliminary remarks are made on their experimental implications. In Sec.~\ref{sec:methods}, the theory in support of the results is systematically presented, followed by a complementary energy conversion analysis in Sec.~\ref{sec:energy}. We discuss the results and conclude the paper in Sec.~\ref{sec:discussions}. Some calculations of technical interest are displayed in the appendices A, B and C.  

\section{Results}
\textit{System}. We consider ballistic SPWs in a metal film surrounded by vacuum. By the so-called jellium model~\cite{pines,abrikosov}, the metal is described as a free electron gas embedded in a static background of homogeneously distributed positive charges. This description is valid if the length scale in question is much longer than the microscopic lattice constant and inter-band transitions are negligible. The kinetic energy of electrons is $\varepsilon(\mathbf{v}) = \frac{1}{2}m\mathbf{v}^2$, where $m$ and $\mathbf{v}$ denote the mass and velocity of the electrons, respectively. The film resides in the region $0\leq z\leq d$ with two surfaces located at $z=0$ and $z=d$, respectively. The surfaces are treated as geometric planes of a hard wall type and they strictly prevent electrons from leaking out of the metal. To simplify our analysis, the surfaces are assumed with identical properties so that the system is symmetric about the mid-plane $z=d/2$. To avoid quantum size effects, we assume $d\gg \hbar/mv_F$, where $\hbar$ is the reduced Planck constant. Throughout we write $\mathbf{x}=(\mathbf{r},z)$ and reserve $\mathbf{r}=(x,y)$ for planar components while let $t$ be the time. We neglect retardation effects in total~\cite{fetter1986,deng2015}. 

\begin{figure*}
\begin{center}
\includegraphics[width=0.97\textwidth]{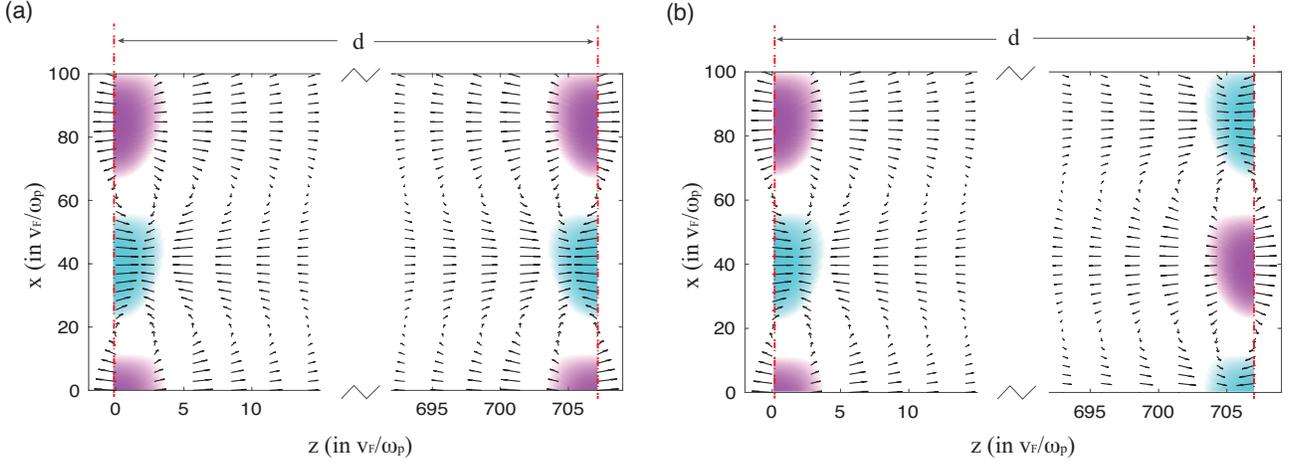}
\end{center}
\caption{Snapshots of the charge density (color) and electric field (arrows) of SPWs supported in a metal film in the region $z\in [0,d]$. $k/k_s=0.1$, $d=500/k_s$ and $p=1$, with $k_s = (\omega_p/\sqrt{2})/v_F$. The symmetric mode $\rho_+(z)$ and anti-symmetric mode $\rho_-(z)$ are displayed in panels (a) and (b), respectively. \label{figure:f1}}
\end{figure*}

\textit{Results}. With two surfaces, a film possesses two branches of SPWs, which at large $d$ degrade into those for two semi-infinite metals. Reflection symmetry about the mid-plane requires the corresponding charge densities to bear a definite sign under the reflection. The branch whose charge density is invariant under the reflection is called symmetric while the one whose charge density changes sign under reflection is called anti-symmetric. In the literature, the symmetric and anti-symmetric SPWs are also designated as \textit{tangential} and \textit{normal} oscillations, respectively. Profiles of the charge densities for symmetric and anti-symmetric SPWs are mapped in Fig.~\ref{figure:f1} (a) and (b), respectively, together with the electric field accompanying them. 

\begin{figure*}
\begin{center}
\includegraphics[width=0.97\textwidth]{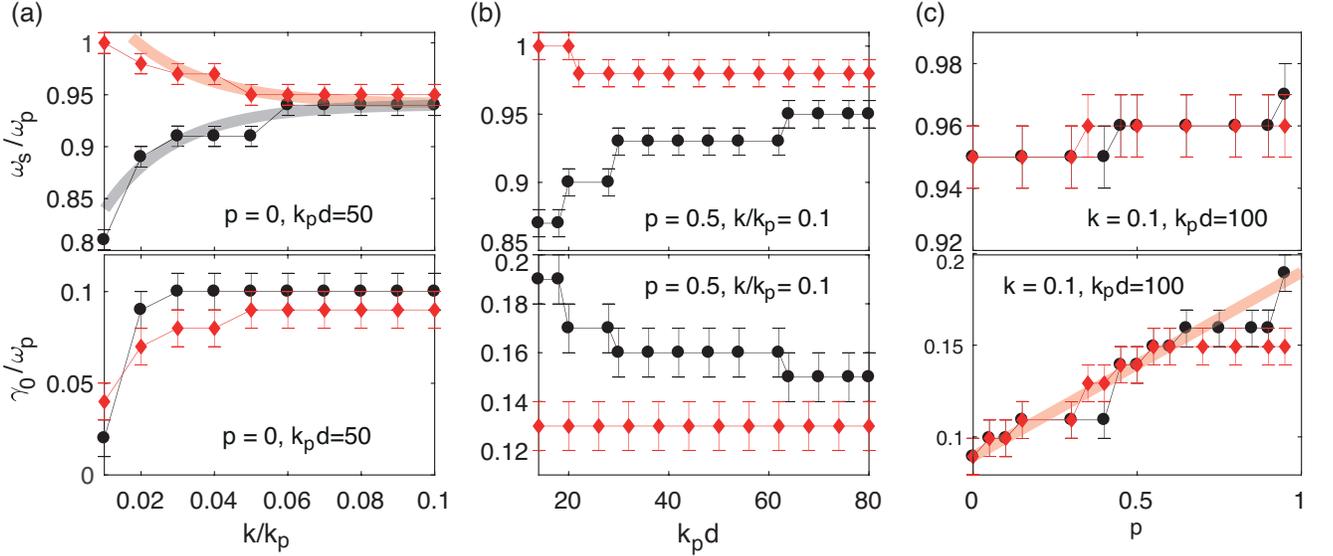}
\end{center}
\caption{Plot of the SPW frequency $\omega_s$ and self-amplification rate $\gamma_0$ versus wavenumber $k$, film thickness $d$ and surface roughness parameter $p$. Circles and diamonds are designated for symmetric and anti-symmetric modes, respectively. $\omega_s$ and $\gamma_0$ are obtained by numerically solving Eq.~(\ref{51}), with Landau damping automatically included. $k_p = \omega_p/v_F$. The cut-off $q_c=1.5k_p$ has been used. The error bar is $\pm 0.01$, corresponding to the grid resolution of $\bar{\omega}$ in the complex frequency plane used in our numerical method. In the upper panel of (a), the thick lines are given by $\propto \sqrt{1-(1\mp e^{-kd})(1+p)/4}$ with $p=0$. In the lower panel of (c), the thick line is $\sim 0.1 \times (1+p)$. \label{figure:f2}}
\end{figure*}

We find that the SPW frequency $\omega^{\pm}_s$ is significantly (as much as $30$\%) higher than $\omega^{\pm}_{s0} = \left(\omega_p/\sqrt{2}\right)\sqrt{1\pm e^{-kd}}$ which would be obtained by the hydrodynamic/Drude theory. Here the plus (minus) sign is affixed and refers to symmetric (anti-symmetric) modes, $\omega_p$ denotes the characteristic plasma frequency of the metal and $k$ is the SPW wavenumber. The dependences of $\omega^{\pm}_s$ on $k$, $d$ and surface scattering -- the effects of which could be summarized in the \textit{Fuchs} parameter $p$ in the simplest possible scattering picture, are displayed in the upper panels of Fig.~\ref{figure:f2} (a), (b) and (c), respectively. The great contrast between $\omega^{\pm}_s$ and $\omega^{\pm}_{s0}$ would be ideal for experimentally verifying our theory. Unfortunately, in the most commonly experimented materials, such as noble metals, due to pronounced inter-band transitions there is no simple relation between $\omega_p$ and $\omega^{\pm}_{s0}$. 

More interestingly, we reveal a universal self-amplification channel for SPWs irrespective of their symmetry. Namely, we find that the net amplification rate of SPWs can be generally written as $\gamma^{\pm} = \gamma^{\pm}_0 - \tau^{-1}$, where $\gamma_0$ is warranted to be non-negative by a general principle and independent of $\tau$. In the conventional theory, $\gamma^{\pm}_0$ vanishes identically and amplification would be impossible without extrinsic energy supply~\cite{bergman2003,seidel2005,leon2008,leon2010,yu2011,pierre2012,fedyanin2012,cohen2013}. The dependences of $\gamma^{\pm}_0$ on $k$, $d$ and $p$ are shown in the lower panels of Fig.~\ref{figure:f2} (a), (b) and (c), respectively, where we observe that (1) $\gamma^{\pm}_0$ is generally a sizable fraction (as much as $\sim 10$\%) of $\omega_p$, (2) it increases as $k$ increases, i.e. higher amplification obtains for shorter wavelengths and (3) it increases as $p$ increases, i.e. smooth surfaces produce higher amplification than rough surfaces. We also see that $\gamma^{+}_0$ is more sensitive to film thickness than $\gamma^{-}_0$. 

\begin{figure*}
\begin{center}
\includegraphics[width=0.97\textwidth]{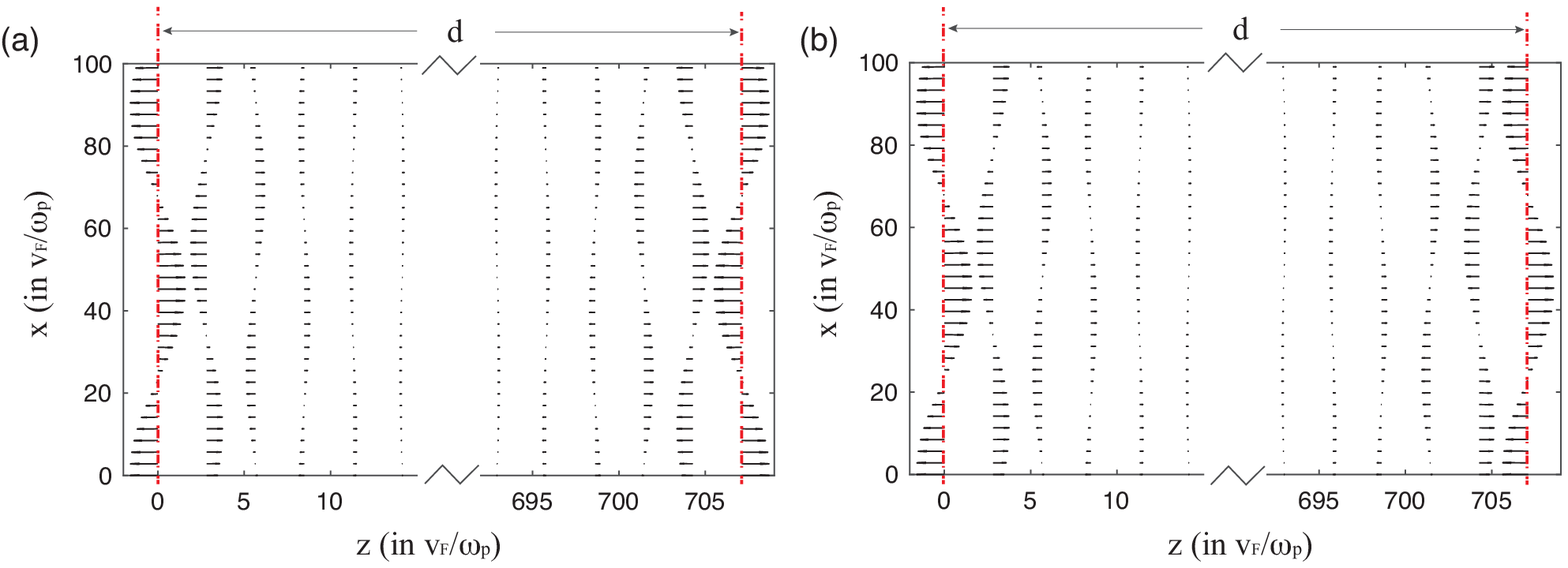}
\end{center}
\caption{Snapshots of the surface-ballistic current density for (a) symmetric and (b) anti-symmetric modes, with the same parameters as in Fig.~\ref{figure:f2}. Note that the currents are directed largely normal to the surface, unlike diffusive currents, which are largely normal to the local electric field. \label{figure:f3}}
\end{figure*}

Additionally, we show that the electrical current density in the system can be split into two disparate components, which we call $\mathbf{J}_D$ and $\mathbf{J}_B$, respectively. An example of their profiles is exhibited in Fig.~\ref{figure:f3} (a) and (b) respectively for the symmetric and anti-symmetric modes. What critically sets them apart rests with their distinct relations with the electric field $\mathbf{E}$ present in the system. $\mathbf{J}_D$ responds to $\mathbf{E}$ as if the system had no surfaces and is therefore primarily a bulk property. As such, it can also be satisfactorily captured by the hydrodynamic/Drude model. For this reason, we designate it a diffusive component, regardless of the value of $\tau$. On the contrary, $\mathbf{J}_B$ represents genuine surface effects and would totally disappear were the surfaces absent. In particular, it synthesizes the effects ensuing from the fact that the system is not translationally invariant along the direction normal to the surfaces. These effects are completely beyond the hydrodynamic/Drude model but well within the scope of Boltzmann's approach, which is employed in our theory to be expounded in the next section. We thus designate $\mathbf{J}_B$ as a surface-ballistic component.  

Finally, we find that the self-amplification channel is a direct consequence of $\mathbf{J}_B$. Indeed, were not for $\mathbf{J}_B$, SPWs would behave in accord with the hydrodynamic/Drude model. This is already clear from the orientations of $\mathbf{J}_{D/B}$ relative to $\mathbf{E}$. As seen in Fig.~\ref{figure:f3}, $\mathbf{J}_D$ points at right angles with $\mathbf{E}$ almost locally, whereas $\mathbf{J}_B$ flows normal to the surface paying little regard to $\mathbf{E}$. Therefore, $\mathbf{E}$ does no work on $\mathbf{J}_D$ on average while, as shown in Sec.~\ref{sec:energy}, it does a negative amount of work on $\mathbf{J}_B$, thereby imparting energy from the electrons to SPWs and destabilizing the Fermi sea. 

\textit{Remarks}. Experimentally verifying the self-amplification channel and the theory in general would be of considerable interest, as it would drastically change the way we conceive and utilize SPWs and renew our interest in surface science in a broad sense. The self-amplification channel could manifest itself for instance in the temperature dependence of various spectra, e.g. electron loss spectra. We discuss this aspect in Sec.~\ref{sec:discussions}. Here we mainly concern ourselves with the experimental implications of the surface-ballistic current $\mathbf{J}_B$. 

Being an integral part of the electrical responses of metals, $\mathbf{J}_B$ is expected to play a role in virtually every phenomena where surface is not negligible. Examples include electron energy losses, reflectance and van der Waals forces. Unlike $\mathbf{J}_D$, which does not reflect surface scattering effects, $\mathbf{J}_B$ is surface-specific via the \textit{Fuchs} parameter. Moreover, they differ in phase by $\sim \pi/2$. To be specific, let us consider the Ferrel radiation~\cite{ferrell1958}. Ferrel predicted that anti-symmetric SPWs in thin films would radiate in a characteristic pattern. Some experiments even claimed to have observed this radiation~\cite{brown1960,arakawa1964,donohue1986}. Ferrel considered only $\mathbf{J}_D$. Following him, we find that including $\mathbf{J}_B$ could boost the radiation power by a factor $\sim 1+(3/2\pi)^2(1+2p)^2$. Though a crude estimate, it does imply that surface properties could be utilized to tune the radiation. In this paper, we focus on the fundamental theory of ballistic SPWs. A systematic treatment of Ferrel radiation will be published elsewhere. 

As aforementioned, a major obstacle in experimentally studying the theory lies with inter-band transitions, which have been neglected in our theory. A detailed discussion of their effects is presented in Sec.~\ref{sec:discussions}. 

\section{Theory}
\label{sec:methods}
This section is devoted to a thorough exposition of the theory. We begin with a discussion of the equation of continuity in the presence of surfaces. Thence we proceed to Boltzmann's approach and analyze how to handle surface effects in this approach. The electronic distribution functions, obtained by solving Boltzmann's equation, are discussed in detail. The electrical current densities are then calculated and the exact equation of motion for the charge density is established. Solutions to the equation are discussed and the properties of SPWs are analyzed. Various limits are presented and connections are made with the hydrodynamic/Drude models. 

\subsection{Equation of Continuity}
\label{sub:21}
The starting point of our theory is the equation of continuity, $\left(\partial_t+1/\tau\right)\rho(\mathbf{x},t)+\partial_{\mathbf{x}}\cdot\mathbf{j}(\mathbf{x},t)=0$, which relates the charge density $\rho(\mathbf{x},t)$ and the current density $\mathbf{j}(\mathbf{x},t)$ in a universal manner. Here $\mathbf{j}(\mathbf{x},t)$ arises in the presence of an electric field $\mathbf{E}(\mathbf{x},t)$ and the damping term $-\rho(\mathbf{x},t)/\tau$ is included to account for the thermal currents due to electronic collisions that would drive the system toward thermodynamic equilibrium. In the jellium model, $\rho(\mathbf{x},t)$ appears when the electron density is perturbed away from its equilibrium value $n_0$. 

As the surfaces strictly prevent electrons from escaping the metal, we may write $\mathbf{j}(\mathbf{x},t) =\left[\Theta(z)-\Theta(z-d)\right]\mathbf{J}(\mathbf{x},t)$, where $\Theta(z)$ is the Heaviside step function. In doing this, we have embodied the surfaces as hard walls and considered the fact that $\mathbf{J}(\mathbf{x},t)$ may not vanish even in the immediate neighborhood of the surfaces -- as is obviously the case with Drude model. With this prescription, the equation of continuity can be rewritten
\begin{equation}
\left(\partial_t+\frac{1}{\tau}\right)\rho(\mathbf{x},t)+\partial_{\mathbf{x}}\cdot\mathbf{J}(\mathbf{x},t) = S(\mathbf{x},t), \label{1}
\end{equation} 
where the effective source term
\begin{equation}
S(\mathbf{x},t) = J_z(\mathbf{x}_d,t)\delta(z-d)-J_z(\mathbf{x}_0,t)\delta(z),
\label{2}
\end{equation}
results directly from the presence of the surfaces. Here $\mathbf{x}_0 = (\mathbf{r},0)$ and $\mathbf{x}_d = (\mathbf{r},d)$ denote points on the surface at $z=0$ and those on that at $z=d$, respectively. Physically, $S(\mathbf{x},t)$ corresponds to the scenario that charges must pile up on the surfaces if they do not come to a halt before they reach them. 

Without loss of generality we seek fields in this form: $\rho(\mathbf{x},t) = $ Re $\left[\rho(z)e^{i(kx-\omega t)}\right]$ and $\mathbf{J}(\mathbf{x},t) = $ Re $\left[\mathbf{J}(z)e^{i(kx-\omega t)}\right]$. Similarly, for the electric field $\mathbf{E}(\mathbf{x},t) = $ Re $\left[\mathbf{E}(z)e^{i(kx-\omega t)}\right]$ and the electrostatic potential $\phi(\mathbf{x},t) = $ Re $\left[\phi(z)e^{i(kx-\omega t)}\right]$. In these expressions, Re/Im takes the real/imaginary part of a quantity,  $k\geq0$ is a wavenumber and $\omega$ is the eigen-frequency to be determined. Equation~(\ref{1}) becomes 
\begin{equation}
-i\bar{\omega}\rho(z) + \mathbf{\nabla}\cdot\mathbf{J}(z) = S(z), \label{1b}
\end{equation}
where $\bar{\omega} = \omega+i/\tau$, $\mathbf{\nabla} = (ik,\partial_y,\partial_z)$ and 
\begin{equation}
S(z) = J_z(d)\delta(z-d)-J_z(0)\delta(z).
\label{source}
\end{equation}
Equation (\ref{1b}) will serve as the equation of motion for $\rho(z)$ when supplemented with additional relations to be formulated between $\mathbf{J}(z)$ and $\rho(z)$ in what follows.    

\subsection{The Law of Electrostatics}
\label{sub:22}
If the SPW phase velocity is much smaller than the speed of light $c$ in vacuum, i.e. $k>k_0$, where $k_0$ is the wavenumber of light at the SPW frequency, the system will be in the non-retarded regime~\cite{deng2015} and we can relate $\phi(\mathbf{x},t)$ and $\rho(\mathbf{x},t)$ by the laws of electrostatics. Without external charges, we have~\cite{deng2015} $$\phi(z) = \frac{2\pi}{k}~\int dz'~e^{-k~\abs{z-z'}}\rho(z').$$ Instead of $\rho(z)$, we directly work with its Fourier components. Generically, we may write $$\rho(z) = \sum^{\infty}_{n=0} \rho_n\cos(q_nz), ~q_n = \frac{\pi n}{d}.$$ The components are given by
\begin{eqnarray}
\rho_n = \frac{1}{d_n}\int^d_0dz~ \rho(z)\cos(q_nz), ~d_n = \frac{d}{2-\delta_{n,0}}, \label{9}
\end{eqnarray}
where $\delta_{m,n}$ denotes the Kroneker symbol. 

As the surfaces of the film are assumed identical, the system is invariant under reflection about its mid-plane. This symmetry makes it useful to write $\rho(z)$ as a superposition of a symmetric mode $\rho_+(z)$ and an anti-symmetric mode $\rho_{-}(z)$. Namely, 
$$\rho(z) = \rho_+(z) + \rho_-(z),$$
where $\rho_{+}(z)$ includes all the terms with even $n$ whereas $\rho_{-}(z)$ those with odd $n$. As such, $\rho_+(0) = \rho_+(d)$ and $\rho_-(0)=-~\rho_-(d)$. Due to the symmetry $\rho_+(z)$ and $\rho_-(z)$ will be shown to be strictly decoupled. We impose on $q_n$ a cutoff $q_c$ of the order of a reciprocal lattice constant; otherwise, the jellium model would cease to be valid. Obviously, $q_c\sim k_F$, where $k_F$ is the Fermi wavenumber of the electrons in the metal. 

In terms of $\rho_n$, we can rewrite
\begin{equation}
\phi(z) = \sum^{\infty}_{n=0}\frac{2\pi \rho_n}{k^2+q^2_n}\left[2\cos(q_nz)-e^{-kz}-(-1)^ne^{-k(d-z)}\right], \label{10}
\end{equation}
The electric field, $\mathbf{E}(z) = -~\mathbf{\nabla}\phi(z)$, can then be obtained straightforwardly. In equation (\ref{10}) the exponentials, $e^{-kz}$ and $e^{-k(d-z)}$, would all vanish if the surfaces were sent to infinity. We may then write $\phi(z) = \phi^{bulk}(z) + \phi^{surface}(z)$, where $\phi^{surface}(z)$ includes the contributions from all the exponentials while $\phi^{bulk}(z)$ contains the remaining contributions. Accordingly, $\mathbf{E}(z) = \mathbf{E}^{bulk}(z) + \mathbf{E}^{surface}(z)$. Such a partition proves useful in analyzing surface specific effects. 

\subsection{Electronic Distribution Function}
\label{sub:23}
The electric field $\mathbf{E}(\mathbf{x},t)$ drives an electrical current $\mathbf{J}(\mathbf{x},t)$. We employ Boltzmann's equation, which is valid as long as inter-band transitions are negligible, to calculate this current. Including the transitions in our formalism is straightforward but will be skipped here. Surfaces scatter electrons. On the microscopic level, one can in principle introduce a surface potential $\phi_s(\mathbf{x})$ in Boltzmann's equation to produce such scattering. The corresponding surface field $\mathbf{E}_s(\mathbf{x}) = -\partial_{\mathbf{x}}\phi_s(\mathbf{x})$ should be peaked on the surfaces and may have an infinitesimal spread complying with the hard-wall picture of surfaces. However, as $\phi_s(\mathbf{x})$ can hardly be known and varies from one sample to another, this method is impractical and futile. 

Alternatively surface scattering effects can be dealt with using boundary conditions. This is possible because $\mathbf{E}_s(\mathbf{x})$ acts only within the immediate neighborhoods of the surfaces; In the bulk of the sample, the electronic distribution function $f(\mathbf{x},\mathbf{v},t)$ sought as solutions to Boltzmann's equation can be specified up to some parameters, which summarize the effects of -- while without actually knowing -- $\phi_s(\mathbf{x})$. With translational symmetry along the surfaces, only one such parameter, i.e. the so-called \textit{Fuchs} parameter $p$, is needed in the simplest model. Physically, $p$ measures the probability that an electron is bounced back when impinging upon the surface. We write $f(\mathbf{x},\mathbf{v},t) = f_0(\varepsilon(\mathbf{v}))+g(\mathbf{x},\mathbf{v},t)$, where $f_0(\varepsilon)$ denotes the Fermi-Dirac distribution and $g(\mathbf{x},\mathbf{v},t)$ represents the non-equilibrium part due to the presence of $\mathbf{E}(\mathbf{x},t)$. The current density can then be calculated by $\mathbf{J}(\mathbf{x},t) = (m/2\pi\hbar)^2\int d^3\mathbf{v}~e\mathbf{v}~g(\mathbf{x},\mathbf{v},t)$, where $e$ denotes the charge of an electron. It is worth pointing out that, as $g(\mathbf{x},\mathbf{v},t)$ is a distribution for the bulk, the actual charge density is not given by $\tilde{\rho}(\mathbf{x},t) = (m/2\pi\hbar)^2\int d^3\mathbf{v}~e~g(\mathbf{x},\mathbf{v},t)$, i.e. $\rho(\mathbf{x},t)\neq\tilde{\rho}(\mathbf{x},t)$. Actually, $\tilde{\rho}(\mathbf{x},t)$ satisfies $(\partial_t+1/\tau)\tilde{\rho}(\mathbf{x},t)+\partial_{\mathbf{x}}\cdot\mathbf{J}(\mathbf{x},t)=0$ rather than Eq.~(\ref{1}). By comparison, one sees that what is missing from $\tilde{\rho}(\mathbf{x},t)$ is the charges localized on the surface. 

As before we write $g(\mathbf{x},\mathbf{v},t) = $ Re $\left[g(\mathbf{v},z)e^{i(kx-\omega t)}\right]$. For linear responses, Boltzmann's equation can be written
\begin{equation}
\frac{\partial g(\mathbf{v},z)}{\partial z} + \lambda^{-1}~g(\mathbf{v},z) +
 ef'_0(\varepsilon)~\frac{\mathbf{v}\cdot\mathbf{E}(z)}{v_z} = 0,  
\label{11}
\end{equation} 
where $ \lambda = iv_z/\tilde{\omega}$ with $\tilde{\omega} = \bar{\omega}-kv_x$ and $f'_0(\varepsilon) = \partial f_0/\partial \varepsilon(\mathbf{v})$. In this equation, the velocity $\mathbf{v}$ is more of a parameter than an argument and can be used to tag electron beams. It is straightforward to solve the equation under appropriate boundary conditions~\cite{see Appendix B}. We divide $g(\mathbf{v},z)$ into a bulk and a surface term, i.e. $$g(\mathbf{v},z) = g_{bulk}(\mathbf{v},z)+g_{surface}(\mathbf{v},z),$$ where the bulk term would exist even in the absence of surfaces whereas the surface term would not. Using Eq.~(\ref{10}) for $\mathbf{E}(z)$, we obtain
\begin{equation}
g_{bulk}(\mathbf{v},z) = -ef'_0\sum^{\infty}_{n=-\infty}\frac{2\pi\rho_n}{k^2+q^2_n}
\frac{kv_x+q_nv_z}{\bar{\omega}-(kv_x+q_nv_z)}e^{iq_nz}, \label{13}
\end{equation}
where we have defined $\rho_{n<0} := \rho_{-n}$. For large $d$ equation (\ref{13}) converges to the distribution function of a boundless system for either the symmetric mode or the anti-symmetric mode. It is notable that $g_{bulk}(\mathbf{v},z)$ bears a single form for all electrons regardless of their velocities. 

As for $g_{surface}(\mathbf{v},z)$, we find it with a subtle structure: it can be written as a sum of two contributions, one of which, $g_{D,surface}(\mathbf{v},z)$, has a single form for all electrons irrespective of their velocities while the other, $g_{B,surface}(\mathbf{v},z)$, does not. Explicitly, we find $$g_{D,surface}(\mathbf{v},z) = g^{(1)}_{D,surface}(\mathbf{v},z)+g^{(2)}_{D,surface}(\mathbf{v},z),$$ where
\begin{eqnarray}
g^{(1)}_{D,surface}(\mathbf{v},z) =-ef'_0\sum^{\infty}_{n=0}\frac{2\pi\rho_n}{k^2+q^2_n}\frac{k(v_z-iv_x)}{kv_z+i\tilde{\omega}}e^{-kz}, 
\end{eqnarray}
and 
\begin{eqnarray}
g^{(2)}_{D,surface}(\mathbf{v},z) = -ef'_0\sum^{\infty}_{n=0}\frac{2\pi\rho_n(-1)^n}{k^2+q^2_n}\frac{k(v_z+iv_x)}{kv_z-i\tilde{\omega}}e^{-k(d-z)}. 
\end{eqnarray}
originate from the surfaces at $z=0$ and $z=d$, respectively. 

We may combine $g_{bulk}(\mathbf{v})$ and $g_{D,surface}(\mathbf{v},z)$ in a single term, $$g_D(\mathbf{v},z) = g_{bulk}(\mathbf{v},z) + g_{D,surface}(\mathbf{v},z),$$ in order to separate them from $$g_B(\mathbf{v},z):= g_{B,surface}(\mathbf{v},z).$$ The subscripts, $D$ and $B$, refer to 'diffusive' and 'surface-ballistic', respectively. In so doing, we have decomposed $$g(\mathbf{v},z) = g_D(\mathbf{v},z)+g_B(\mathbf{v},z)$$ in a diffusive and a surface-ballistic component. It is underlined that $g_B(\mathbf{v},z)$ arises only when the surfaces are present. For boundless systems without surfaces, it does not exist even if the electronic motions are totally ballistic, i.e. $\tau\rightarrow\infty$. In other words, $g_B(\mathbf{v},z)$ represents genuine surface effects. It may be interpreted as a contribution from electrons which experience the electric field only on the surfaces and propagate freely in the body. Its expressions are given in what follows.  

\begin{widetext}
Electrons in the film can bounce back and forth between its surfaces. Each bounce gives a factor $p_1p_2e^{2i\tilde{\omega}d/\abs{v_z}}$, whose magnitude is generally smaller than unity (see Appendix~\ref{sec:edf}). Here $p_1$ and $p_2$ are the \textit{Fuchs} parameters for the surfaces at $z=0$ and $z=d$, respectively. Consequently, we neglect multiple bounces, which allows to write $$g_{B,surface}(\mathbf{v},z) = g^{(1)}_{B,surface}(\mathbf{v},z)+ g^{(2)}_{B,surface}(\mathbf{v},z),$$ where $g^{(1)}_{B,surface}(\mathbf{v},z)$ and $g^{(2)}_{B,surface}(\mathbf{v},z)$ originate from the surfaces at $z=0$ and $z=d$, respectively. They are given by 
\begin{eqnarray}
g^{(1)}_{B,surface}(\mathbf{v},z) = \Theta(v_z)~e^{i\frac{\tilde{\omega}z}{v_z}}\left[g^{(1)}_{B,emg}(\mathbf{v})+p_1~g^{(1)}_{B,ref}(\mathbf{v})\right], \quad g^{(2)}_{B,surface}(\mathbf{v},z) = \Theta(-v_z)~e^{i\frac{\tilde{\omega}(z-d)}{v_z}}\left[g^{(2)}_{B,emg}(\mathbf{v})+p_2~g^{(2)}_{B,ref}(\mathbf{v})\right], \nonumber
\end{eqnarray}
where $g^{(1/2)}_{B,emg}(\mathbf{v},z)$ is contributed by electrons that directly emerge from the surface at $z=0/d$ while $g^{(1/2)}_{B,ref}(\mathbf{v},z)$ by reflected electrons and hence proportional to $p_{1/2}$. In what follows we take $p_1 = p_2 = p$. The expressions of $g^{(1/2)}_{B,emg/ref}(\mathbf{v})$ are involved but with a recognizable structure:
\begin{equation}
g^{(1/2)}_{B,emg}(\mathbf{v}) = ef'_0\sum^{\infty}_{n=0}\frac{2\pi\rho_n\langle1|(-1)^n\rangle}
{k^2+q^2_n}\left[\frac{2(\tilde{\omega}kv_x)+q^2_nv^2_z}{\tilde{\omega}^2-q^2_nv^2_z}+\frac{k(v_z\mp iv_x)}{kv_z\pm i\tilde{\omega}}+(-1)^ne^{-kd}\frac{k(v_z\pm iv_x)}{kv_z \mp i\tilde{\omega}}\right], \label{18}
\end{equation}
where the symbol $\langle 1|(-1)^n\rangle$ returns $1$ and $(-1)^n$ for $g^{(1)}_{B,emg}$ and $g^{(2)}_{B,emg}$, respectively. In addition, we have
\begin{eqnarray}
g^{(1)}_{B,ref}(\mathbf{v}) = ef'_0\sum^{\infty}_{n=0} \frac{2\pi\rho_n}{k^2+q^2_n}\left[\left(e^{i\frac{\tilde{\omega}d}{v_z}}(-1)^n-1\right)\frac{2(\tilde{\omega}kv_x-q^2_nv^2_z)}{\tilde{\omega}^2-q^2_nv^2_z}+\left(1-e^{\left(\frac{i\tilde{\omega}}{v_z}-k\right)d}\right)\frac{k(v_z-iv_x)}{kv_z-i\tilde{\omega}}+(-1)^n\left(e^{-kd}-e^{i\frac{\tilde{\omega}d}{v_z}}\right)\frac{k(v_z+iv_x)}{kv_z+i\tilde{\omega}}\right], 
\end{eqnarray}
and
\begin{eqnarray}
g^{(2)}_{B,ref}(\mathbf{v}) = ef'_0\sum^{\infty}_{n=0} \frac{2\pi\rho_n(-1)^n}{k^2+q^2_n}\left[\left(e^{-i\frac{\tilde{\omega}d}{v_z}}(-1)^n-1\right)\frac{2(\tilde{\omega}kv_x-q^2_nv^2_z)}{\tilde{\omega}^2-q^2_nv^2_z}+\left(1-e^{-\left(\frac{i\tilde{\omega}}{v_z}+k\right)d}\right)\frac{k(v_z+iv_x)}{kv_z+i\tilde{\omega}}+(-1)^n\left(e^{-kd}-e^{-i\frac{\tilde{\omega}d}{v_z}}\right)\frac{k(v_z-iv_x)}{kv_z-i\tilde{\omega}}\right],\label{20}\nonumber\\
\end{eqnarray}

\textit{Positiveness of $\mbox{Im}(\bar{\omega})$}. What sets $g_B(\mathbf{v},z)$ apart from its diffusive counterpart rests with its disparate $z$ dependence. Let us take the contribution originating from the surface at $z=0$ for example. Here $g^{(1)}_{B,surface}(\mathbf{v},z) \propto e^{i\tilde{\omega}z/v_z}\propto e^{-\mbox{Im}(\bar{\omega})z/v_z}$, where $v_z\geq 0$. Unless Im$(\bar{\omega})\geq 0$, this expression would diverge for small $v_z$. As such, we may conclude that Im$(\bar{\omega})\geq 0$, a result to be confirmed in what follows by specific calculations. In Appendix~\ref{sec:edf}, we frame this result as a consequence of the causality principle: out-going electrons are determined by in-coming ones; not otherwise.  
\end{widetext}

\subsection{Current Densities}
\label{sub:24}
We are now prepared to discuss the behaviors of the current density, which is written $\mathbf{J}(z) = \mathbf{J}_D(z) + \mathbf{J}_B(z)$, where $$\mathbf{J}_{D/B}(z) = (m/2\pi\hbar)^3\int d^3\mathbf{v}~e\mathbf{v}~g_{D/B}(\mathbf{v},z)$$ is the diffusive/surface-ballistic component of $\mathbf{J}(z)$. The equation of motion for $\rho(z)$ follows upon inserting $\mathbf{J}(z)$ in Eq.~(\ref{1b}). In our calculations, the zero temperature is assumed whenever a concrete form of $f_0(\varepsilon)$ is required, though generalization to finite temperatures is straightforward. 

\subsubsection{Diffusive current density}
Since $g_D(\mathbf{v},z)$ consists of a bulk and a surface component, we accordingly write $\mathbf{J}_D(z) = \mathbf{J}_{bulk}(z)+\mathbf{J}_{D,surface}(z)$, where $\mathbf{J}_{bulk}(z)$ and $\mathbf{J}_{D,surface}(z)$ arise from $g_{bulk}(\mathbf{v},z)$ and $g_{D,surface}(\mathbf{v},z)$, respectively. By straightforward manipulation, one may show that $\mathbf{J}_{D,surface}(z)\propto \mathbf{E}^{surface}(z)$. To the lowest order in $kv_F/\omega_p$, where $\omega_p = \sqrt{4\pi n_0e^2/m}$ is the characteristic plasma frequency of the metal, we have
\begin{equation}
\mathbf{J}_{D,surface} = \frac{i}{\bar{\omega}}\frac{\omega^2_p}{4\pi}~\mathbf{E}^{surface}(z), 
\end{equation}
where the pre-factor heading $\mathbf{E}^{surface}(z)$ is recognized as the Drude conductivity. In addition, we find
\begin{equation}
\mathbf{J}_{bulk}(z) = \frac{i}{\bar{\omega}}\frac{\omega^2_p}{4\pi}\mathbf{E}^{bulk}(z) + \mathbf{J}'(z), 
\end{equation}
where 
\begin{equation}
\mathbf{J}'(z) = \sum^{\infty}_{n=-\infty}\frac{2\pi \rho_n e^{iq_nz}}{k^2+q^2_n} ~\mathbf{F}(k,q_n;\bar{\omega}).
\end{equation}
signifies non-local electrical responses that would engender dispersive plasma waves. In the expression
\begin{equation}
\mathbf{F}(k,q;\bar{\omega}) = \left(\frac{m}{2\pi\hbar}\right)^3
 \int d^3\mathbf{v} (-e^2f'_0)~\mathbf{v}\sum^{\infty}_{l=2}\left(\frac{kv_x+qv_z}{\bar{\omega}}\right)^l. \label{F}
\end{equation}
Only terms with odd $l$ contribute in the series. Note that the normal component of $\mathbf{J}'(z)$ vanishes identically at all surfaces, i.e. $J'_{z}(0) = J'_{z}(d) = 0.$ 

Piecing everything together we obtain $$\mathbf{J}_{D}(z) = \frac{i}{\bar{\omega}}\frac{\omega^2_p}{4\pi}\mathbf{E}(z) + \mathbf{J}'(z).$$ As in the hydrodynamic/Drude model, which is valid only for diffusive electronic motions, the relation between $\mathbf{J}_D(z)$ and $\mathbf{E}(z)$ assumes the form of a generalized Ohm's law. This is why we consider $\mathbf{J}_D(z)$ a diffusive component, irrespective of the value of $\tau$. Its divergence is easily found to be 
\begin{eqnarray}
\label{24}
\mathbf{\nabla}\cdot \mathbf{J}_D(z) =  \frac{i}{\bar{\omega}}\sum^{\infty}_{n=0}\Omega^2(k,q_n;\bar{\omega})\rho_n\cos(q_nz),
\end{eqnarray}
where, with $\mathbf{k} := (k,q)$, 
\begin{equation}
 \quad \Omega^2(k,q;\bar{\omega}) = \omega^2_p + \frac{4\pi\bar{\omega}~\mathbf{k}\cdot \mathbf{F}(k,q;\bar{\omega})}{\mathbf{k}\cdot\mathbf{k}}.
\end{equation}
Fourier transforming Eq.~(\ref{24}) yields
\begin{equation}
\frac{1}{d_n}\int^{d}_0 dz\cos\left(q_n z\right)~\mathbf{\nabla}\cdot\mathbf{J}_D(z) = \frac{i}{\bar{\omega}}~\Omega^2(k,q_n;\bar{\omega}) ~\rho_n. \label{26}
\end{equation}

We will show that $\Omega(k,q;\bar{\omega})$ is intimately related to the properties of bulk plasma waves. As expected, $\Omega(k,q;\bar{\omega})$ only depends on the length of $\mathbf{k}$, not its direction. This becomes evident by writing $kv_x+qv_z = \mathbf{k}\cdot\mathbf{v}$ in Eq.~(\ref{F}). The first non-vanishing contribution to $\Omega(k,q;\bar{\omega})$ comes from the term $l=1$ in the series in $\mathbf{F}(k,q;\bar{\omega})$. Retaining only this term, we get 
\begin{equation}
\Omega^2(k,q;\bar{\omega}) \approx \omega^2_p~\left[1+\frac{3}{5}\frac{(k^2+q^2)v^2_F}{\bar{\omega}^2}\right].\label{hy}
\end{equation}
Upon replacing $\bar{\omega}$ with $\omega_p$, one immediately revisits the dispersion relation for bulk waves, which could also be reached through the hydrodynamic model. In the Drude model, the dispersion is totally neglected. 

It is noted that $\Omega(k,q;\bar{\omega})$ generally possesses an imaginary part. In case Im$(\bar{\omega})$ is vanishingly small, the imaginary part arises from a pole, located at $\bar{\omega} = kv_x+qv_z$, in the integrand in Eq.~(\ref{F}), giving rise to Landau damping in bulk waves and SPWs. In our numerical computation of $\bar{\omega}$, Landau damping will be automatically included. 

\begin{widetext}
\subsubsection{Surface-ballistic current density} 
Separating the contributions of emerging electrons from that of reflected electrons, we write $\mathbf{J}_B(z) = \mathbf{J}_{B,emg}(z) + p\mathbf{J}_{B,ref}(z)$. Explicitly, we find
\begin{eqnarray}
\mathbf{J}_{B,emg/ref}(z) = \left(\frac{m}{2\pi\hbar}\right)^3\int d^3\mathbf{v} ~e\mathbf{v}~\left[\Theta(v_z)e^{i\frac{\tilde{\omega}z}{v_z}}g^{(1)}_{B,emg/ref}(\mathbf{v})+\Theta(-v_z)e^{i\frac{\tilde{\omega}(z-d)}{v_z}}g^{(2)}_{B,emg/ref}(\mathbf{v})\right] =: \left(\frac{m}{2\pi\hbar}\right)^3\int d^3\mathbf{v} ~ \mathbf{J}_{B,emg/ref}(\mathbf{v},z), 
\end{eqnarray}
where we have defined $\mathbf{J}_{B,emg/ref}(\mathbf{v},z)$ as the contribution from the beam of electrons with velocity $\mathbf{v}$. Using the expressions of $g^{(1/2)}_{B,emg/ref}(\mathbf{v})$ given by Eqs.~(\ref{18}) - (\ref{20}), we can rewrite it
\begin{equation}
J_{B,emg/ref,x|z}(\mathbf{v},z) = \Theta(v_z) ~e^2f'_0~\sum^{\infty}_{n=0}\frac{2\pi\rho_n}{k^2+q^2_n}~L_{emg/ref}\left(v_x,v_z,k,q_n,\bar{\omega},(-1)^n\right)~v_{x|z} \left(e^{i\frac{\tilde{\omega}z}{v_z}}+(+|-)(-1)^ne^{i\frac{\tilde{\omega}(d-z)}{v_z}}\right), \label{re}
\end{equation}
where, with $s=\pm 1$, 
\begin{eqnarray}
L_{emg}(v_x,v_z,k,q,\bar{\omega},s) &=& \frac{2(q^2v^2_z+\tilde{\omega}kv_x)}{\tilde{\omega}^2-q^2v^2_z} + \frac{k(v_z-iv_x)}{kv_z+i\tilde{\omega}} + s~e^{-kd}\frac{k(v_z+iv_x)}{kv_z-i\tilde{\omega}}, \label{34}\\
L_{ref}(v_x,v_z,k,q,\bar{\omega},s) &=& \frac{2(q^2v^2_z-\tilde{\omega}kv_x)}{\tilde{\omega}^2-q^2v^2_z}\left(1-s~e^{i\frac{\tilde{\omega}d}{v_z}}\right) + \frac{k(v_z-iv_x)}{kv_z-i\tilde{\omega}}\left(1-e^{(i\frac{\tilde{\omega}}{v_z}-k)d}\right) + s~e^{-kd}\frac{k(v_z+iv_x)}{kv_z+i\tilde{\omega}}\left(1-e^{(i\frac{\tilde{\omega}}{v_z}+k)d}\right).\label{35}
\end{eqnarray}
In the limit $d\rightarrow\infty$, all the exponentials in $L_{emg/ref}$ vanish and we would recover the result for semi-infinite metals; $\mathbf{J}_{B,emg/ref}(z)$ could then be written as a sum of that for two semi-infinite metals. As expected, the surfaces of the film are decoupled in this limit. For thin films, Eq.~(\ref{re}) implies that $\mathbf{J}_{B,emg/ref}(\mathbf{v},z)$ mainly runs along the surface for symmetric modes while normal to it for anti-symmetric modes. 

The divergence of $\mathbf{J}_B(z)$ can be easily obtained. In the first place we have
\begin{equation}
\mathbf{\nabla}\cdot\mathbf{J}_{B,emg/ref}(z) = i\bar{\omega}\left(\frac{m}{2\pi\hbar}\right)^3\int d^3\mathbf{v}~\Theta(v_z)~e^2f'_0~\sum^{\infty}_{n=0}\frac{2\pi\rho_n}{k^2+q^2_n} L_{emg/ref}(v_x,v_z,k,q_n,\bar{\omega},(-1)^n) ~ \left(e^{i\frac{\tilde{\omega}z}{v_z}}+(-1)^ne^{i\frac{\tilde{\omega}(d-z)}{v_z}}\right), 
\end{equation}
whose Fourier transform is 
\begin{equation}
\frac{1}{d_m}\int^{d}_0dz~\cos(q_mz)~\mathbf{\nabla}\cdot \mathbf{J}_{B,emg/ref}(z) = \frac{i}{\bar{\omega}} \sum^{\infty}_{n=0} \mathcal{M}_{emg/ref,mn}\rho_n, \label{37}
\end{equation}
with 
\begin{equation}
\mathcal{M}_{emg/ref,mn} = \frac{\Gamma_{mn}}{d_m}\frac{2\pi ~\bar{\omega}^2}{k^2+q^2_n}\left(\frac{m}{2\pi\hbar}\right)^3\int d^3\mathbf{v}~\Theta(v_z)~e^2f'_0~\frac{i\tilde{\omega}v_z}{\tilde{\omega}^2-q^2_mv^2_z}\left(1-(-1)^ne^{i\frac{\tilde{\omega}d}{v_z}}\right)~L_{emg/ref}(v_x,v_z,k,q_n,\bar{\omega},(-1)^n).
\end{equation}
Here $\Gamma_{mn} = 1+(-1)^{m+n}$, which would vanish identically unless $m$ and $n$ have the same parity. It follows that
\begin{equation}
\frac{1}{d_m}\int^{d}_0dz~\cos(q_mz)~\mathbf{\nabla}\cdot \mathbf{J}_{B}(z) = \frac{i}{\bar{\omega}} \sum^{\infty}_{n=0} \mathcal{M}_{mn}\rho_n, \quad\mathcal{M}_{mn} = \mathcal{M}_{emg,mn}+p\mathcal{M}_{ref,mn} \label{M}
\end{equation}
We can write $\mathcal{M} = \mathcal{M}^{+}\bigoplus\mathcal{M}^{-}$, where $\mathcal{M}^{\pm} = \mathcal{M}^{\pm}_{emg}+p\mathcal{M}^{\pm}_{ref}$ operates on the space of $\rho_{\pm}(z)$, with
\begin{equation}
\mathcal{M}^{\pm}_{emg/ref,mn} = \frac{1}{d_m}\frac{4\pi ~\bar{\omega}^2}{k^2+q^2_n}\left(\frac{m}{2\pi\hbar}\right)^3\int d^3\mathbf{v}~\Theta(v_z)~e^2f'_0~\frac{i\tilde{\omega}v_z}{\tilde{\omega}^2-q^2_mv^2_z}\left(1\mp e^{i\frac{\tilde{\omega}d}{v_z}}\right)~L_{emg/ref}(v_x,v_z,k,q_n,\bar{\omega},\pm 1). \label{39}
\end{equation}
In Appendix \ref{sec:M}, we show that $\mathcal{M}^{\pm}$ is of the order of $kv_F/\omega_p$.

\subsection{Equation of Motion and SPW Solutions}
\label{sub:26}
\textit{Symmetric and anti-symmetric modes}. We proceed to transform Eq.~(\ref{1b}) into the equation of motion for $\rho(z)$. In the first place let us show that $\rho_+(z)$ and $\rho_-(z)$ are strictly decoupled. As is clear from preceding subsections, $\nabla\cdot\mathbf{J}(z)$ and hence the entire left hand side of Eq.~(\ref{1b}) are block diagonal with respect to the subspaces respectively spanned by $\rho_+(z)$ and $\rho_-(z)$. We can prove that $S(z)$ disconnects the subspaces as well. To this end, we Fourier transform $S(z)$ in Eq.~(\ref{source}) to obtain 
\end{widetext}
$$S_m = \frac{1}{d_m}\int^d_0dz~\cos(q_mz)~S(z) = \frac{1}{d_m}\left[J_z(d)(-1)^m-J_z(0)\right].$$ Linearly depending on $\rho(z) = \rho_+(z)+\rho_-(z)$, $J_z(z)$ can be split as $J_z(z) = J^+_{z}(z) + J^-_{z}(z)$, where $J^{+/-}_{z}(z)$ denotes the contributions from $\rho_{+/-}(z)$. From their expressions given in preceding sessions, we easily deduce that 
\begin{equation}
J^{+/-}_{z}(0) \pm J^{+/-}_{z}(d) = 0, \label{jz}
\end{equation}
by which we rewrite
\begin{equation}
S_m = - \frac{1}{d_m}\left[J^+_{z}(0)\left(1+(-1)^m\right)+J^-_{z}(0)\left(1-(-1)^m\right)\right]. \label{45} 
\end{equation}
This equation allows us to organize $S_m$ in the form of a column vector $\mathcal{S} = \mathcal{S}_+\bigoplus \mathcal{S}_-$, where $\mathcal{S}_{+,l} = S_{2l}$ contains all the elements $m=2l$ with $l=0,1,...$, while $\mathcal{S}_{-,l} = S_{2l+1}$ contains all the elements $m=2l+1$. As such, the symmetric and anti-symmetric modes belong to different sectors and are strictly decoupled. We can write
\begin{equation}
\mathcal{S}_{+/-} = - ~ \frac{4~J^{+/-}_z(0)}{d} ~ \mathbb{E}_{+/-},  \label{46}
\end{equation}
where $\mathbb{E}_{+,l} = 1-\delta_{l,0}/2$ and $\mathbb{E}_{-,l} = 1$. 

\textit{Equation of motion}. The equation of motion is obtained by Fourier transforming Eq.~(\ref{1b}) and using Eqs.~(\ref{26}), (\ref{37}) and (\ref{39}) as well as (\ref{46}). We find 
\begin{equation}
\left[\mathcal{H}^{+/-}(\bar{\omega})- \bar{\omega}^2\mathbb{I}\right]\rho^{+/-} = i\bar{\omega}J^{+/-}_z(0)\frac{4}{d}\mathbb{E}_{+/-}, \label{47}
\end{equation}
where the matrix reads $$\mathcal{H}^{+/-}_{ll'}(\bar{\omega}) = \delta_{l,l'} ~ \Omega^2(k,q^{+/-}_l;\bar{\omega})+\mathcal{M}^{+/-}_{ll'}.$$ Here the column vectors are defined by $\rho^+_{l} = \rho_{2l}$ and $\rho^-_l = \rho_{2l+1}$. We can rewrite
\begin{equation}
J^{+/-}_z(0) = -\frac{i}{\bar{\omega}} \frac{d}{4} \sum^{\infty}_{l=0} \mathcal{G}^{+/-}_{l}~\rho^{+/-}_l = -\frac{i}{\bar{\omega}} \frac{d}{4} ~\mathcal{G}^{+/-} ~ \rho^{+/-}, \label{49}
\end{equation}
where $\mathcal{G}^{+/-} = \mathcal{G}^{+/-}_D + \mathcal{G}^{+/-}_B$ is a row vector. We have
\begin{equation}
\mathcal{G}^{+/-}_{l} = \frac{4\pi~G^{+/-}(k,q^{+/-}_l;\bar{\omega})}{k^2+\left(q^{+/-}_{l}\right)^2} 
\end{equation}
where $G^{+/-}(k,q;\bar{\omega}) = G^{+/-}_D(k)+G^{+/-}_B(k,q;\bar{\omega})$, with
$$G^{+/-}_D(k) = \frac{2}{d}\frac{\omega^2_p}{4\pi} ~k~\left(1\mp e^{-kd}\right),$$ which is comparable to the counterpart for semi-infinite metals, and 
\begin{widetext}
\begin{equation}
G^{+/-}_{B}(k,q;\bar{\omega}) = i\bar{\omega}\frac{2}{d}\left(\frac{m}{2\pi\hbar}\right)^3\int d^3\mathbf{v} \Theta(v_z)(-e^2f'_0)\left(\pm e^{i\frac{\tilde{\omega}d}{v_z}}-1\right)v_z~L(v_x,v_z,k,q,\bar{\omega},\pm 1), \label{gb}
\end{equation}
where $L(v_x,v_z,k,q,\bar{\omega},s) = L_{emg}(v_x,v_z,k,q,\bar{\omega},s)+p~L_{ref}(v_x,v_z,k,q,\bar{\omega},s)$, with $L_{emg/ref}$ given by Eqs.~(\ref{34}) and (\ref{35}). 
\end{widetext}

\textit{SPWs as localized solutions}. Two types of solutions exist to Eq.~(\ref{47}), depending on whether $J^{+/-}_z(0)$ vanishes or not. SPWs are described by solutions with $J^{+/-}_z(0)\neq0$. These solutions represent localized surface waves, for which the equation can be directly solved. We obtain
\begin{equation}
1 = \mathcal{G}^{+/-}~\left[\mathcal{H}^{+/-}(\bar{\omega}) - \bar{\omega}^2\mathbb{I}\right]^{-1}~\mathbb{E}_{+/-},
\end{equation}
which involves no approximations.

Let us write the solution as $\bar{\omega} = \omega_s + i\gamma_0$ and hence the SPW eigen-frequency is given by $\omega = \omega_s+i\gamma$ with $\gamma = \gamma_0-1/\tau$. One can show that $\omega_s+i\gamma_0$ always occurs with $-\omega_s+i\gamma_0$, in accord with the fact that $\rho(\mathbf{x},t)$ is real-valued. We shall take $\omega_s\geq0$ for definiteness. 

Dropping $\mathcal{M}^{+/-}$ as an approximation, the equation becomes
\begin{equation}
1 = \sum^{\infty}_{l=0}\frac{4\pi~G^{+/-}(k,q^{+/-}_l;\bar{\omega})}{k^2+\left(q^{+/-}_{l}\right)^2} ~\frac{(1-\delta_{l,0}/2)~|~1}{\Omega^2(k,q^{+/-}_l;\bar{\omega})-\bar{\omega}^2}. \label{51}
\end{equation}
In addition, we have 
\begin{equation}
\rho^{+/-}_l = \frac{i\bar{\omega}J^{+/-}_z(0)}{\Omega^2(k,q^{+/-}_l;\bar{\omega})-\bar{\omega}^2}\frac{4}{d}\left[(1-\delta_{l,0}/2)~|~1\right]. \label{ch}
\end{equation}

Notably, $\tau$ is not explicitly involved in any of the above equations, implying that the value of $\bar{\omega}$ does not depend on $\tau$. 

\begin{widetext}
\subsection{Approximate and Numerical Solutions}
\textit{Hydrodynamic/Drude limits}. The hydrodynamic model is attained when the surface-ballistic effects, synthesized in the quantity $G^{+/-}_B(k,q;\bar{\omega})$, are ignored in total and the bulk plasma wave dispersion is taken as given by Eq.~(\ref{hy}), i.e. $\Omega(k,q;\bar{\omega})\approx \omega^2_p+(3/5)\left(k^2+q^2\right)v^2_F$. In the Drude model, the dispersion is also ignored. In both models, $\bar{\omega}$ is real-valued and Im$(\omega)=-1/\tau$. Solving Eq.~(\ref{51}) without $G^{+/-}_B(k,q;\bar{\omega})$, for large $d$ we obtain $\omega^{+/-}_s = \omega^{+/-}_{s0}$, with $\omega^{+/-}_{s0} = \left(\omega_p/\sqrt{2}\right)\sqrt{1\pm e^{-kd}}$ for the symmetric/antisymmetric modes of SPWs. Note that the bulk wave frequency always lies above the SPW frequency and hence the factor $1/(\Omega^2(k,q;\omega_p)-\bar{\omega}^2)$ never develops a pole near $\omega_s$: SPWs can not decay via bulk waves.     

\textit{Approximate solutions}. We can solve (\ref{51}) approximately. To the lowest order in $\gamma_0/\omega_s$, we may determine $\omega_s$ by approximating the real part of (\ref{51}) as follows
\begin{eqnarray}
1 \approx \sum^{\infty}_{l=0}\frac{4\pi~\mbox{Re}\left[G^{+/-}\left(k,q^{+/-}_l;\omega_s\right)\right]}{k^2+\left(q^{+/-}_{l}\right)^2} ~ \frac{(1-\delta_{l,0}/2)~|~1}{\Omega^2\left(k,q^{+/-}_l;\omega_s\right)-\omega^2_s}, \label{52}
\end{eqnarray}
The as-obtained $\omega_s$ is then substituted in the imaginary part of Eq.~(\ref{51}) to get $\gamma_0$. We find 
\begin{equation}
\frac{\gamma_0}{\omega_s} \approx - \frac{1}{2}~\frac{\sum^{\infty}_{l=0}\frac{4\pi}{k^2+(q^{+/-}_l)^2}\frac{(1-\delta_{l,0}/2)~|~1}{\Omega^2(k,q^{+/-}_l;\omega_s)-\omega^2_s}\mbox{Im}\left[G^{+/-}(k,q^{+/-}_l;\omega_s)\right]}{\sum^{\infty}_{l=0}\frac{4\pi}{k^2+(q^{+/-}_l)^2}\frac{(1-\delta_{l,0}/2)~|~1}{\Omega^2(k,q^{+/-}_l;\omega_s)-\omega^2_s}\mbox{Re}\left[G^{+/-}(k,q^{+/-}_l;\omega_s)\right]\frac{\omega^2_s}{\Omega^2(k,q^{+/-}_l;\omega_s)-\omega^2_s}}, \label{53}
\end{equation}
which can be brought into a rather simple form if we take $\Omega(k,q;\omega_s)\approx \omega_p$ and $\omega^2_s/\omega^2_p\sim 1/2$ for $kd\gg1$. We get
\begin{equation}
\frac{\gamma_0}{\omega_s} \approx - \frac{1}{2} \frac{\sum^{\infty}_{l=0}\frac{(1-\delta_{l,0}/2)~|~1}{k^2+(q^{+/-}_l)^2}{\mbox{Im}\left[G^{+/-}(k,q^{+/-}_l;\omega_s)\right]}}{\sum^{\infty}_{l=0}\frac{(1-\delta_{l,0}/2)~|~1}{k^2+(q^{+/-}_l)^2}\mbox{Re}\left[G^{+/-}(k,q^{+/-}_l;\omega_s)\right]} = \frac{1}{2}\frac{\mbox{Re}\left[J^{+/-}_z(0)\right]}{\mbox{Im}\left[J^{+/-}_z(0)\right]}, \label{54}
\end{equation}
with $J^{+/-}_z(0)$ evaluated by Eq.~(\ref{49}) with $\omega_s$ in place of $\bar{\omega}$. This relation can also be established by an energy analysis, see Sec.~\ref{sec:energy}. By virtue of the relation that $\mbox{Im}\left[G^{+/-}(k,q;\omega_s)\right] + \mbox{Im}\left[G^{+/-}(k,q;-\omega_s)\right] = 0,$ the same Im$(\bar{\omega})$ exists for $-\omega_s$, as anticipated from the fact that charge density waves are real-valued waves. 

To make progress, we need to evaluate $G^{+/-}_B(k,q;\bar{\omega})$. Writing the integration in Eq.~(\ref{gb}) in spherical coordinates and performing it over the magnitude of $\mathbf{v}$, we arrive at
\begin{eqnarray}
G^{+/-}_B(k,q;\bar{\omega}) = -~i~\frac{2}{d}\frac{\omega^2_p}{4\pi}~\frac{3\bar{\omega}}{4\pi v_F}\int^{2\pi}_0d\varphi \int^{\pi/2}_0 d\theta \sin\theta \cos(\theta) \tilde{L}(v_F\sin\theta\cos\varphi,v_F\cos\theta,k,q,\bar{\omega},\pm), \label{92}
\end{eqnarray}
where we have written $\mathbf{v} = v(\sin\theta\cos\varphi,\sin\theta\sin\varphi,\cos\theta)$ and 
$\tilde{L}(v_x,v_z,k,q,\bar{\omega},s) = \left(1\mp e^{i\tilde{\omega}d/v_z}\right)~L(v_x,v_z,k,q,\bar{\omega},s).$ We expand all factors other than $e^{i\tilde{\omega}d/v_z}$ in $L(v_x,v_z,k,q,\bar{\omega},s)$ into a series of $kv_F/\bar{\omega}$ and retain only the leading term. We find
\begin{eqnarray}
L(v_x,v_z,k,q,\bar{\omega},s) \approx \frac{2q^2v^2_z}{\bar{\omega}^2-q^2v^2_z}\left(1+p-spe^{i\tilde{\omega}d/v_z}\right) + \left(1-se^{-kd}\right)\left[\frac{kv_x}{\bar{\omega}}\left(1-p+pe^{i\tilde{\omega}d/v_z}\right)+\frac{kv_z}{i\bar{\omega}}\left(1-p-pe^{i\tilde{\omega}d/v_z}\right)\right]. \label{L}
\end{eqnarray}
Upon being formally integrated over $\varphi$, the integral in Eq.~(\ref{92}) ends up in this form, 
$\int^{1}_0dr ~ \left(L_0(r)+e^{ir_0/r}L_1(r)+e^{2ir_0/r}L_2(r)\right),$
where only the dependence on $r=\cos\theta$ is explicitly noted down in the integrand and $r_0=\bar{\omega}d/v_F\gg1$. As $e^{ir_0/r}$ and $e^{2ir_0/r}$ are rapidly oscillating functions whereas $L_{1,2,3}(r)$ are slowly varying functions, $L_0(r)$ is the dominant contribution to the integral. We neglect other contributions and obtain
\begin{eqnarray}
G^{+/-}_B(k,q;\bar{\omega}) \approx -~i~\frac{2}{d}\frac{\omega^2_p}{4\pi}\frac{3\bar{\omega}}{v_F}~(1+p)\int^1_0dr~\frac{q^2v^2_Fr^3}{\bar{\omega}^2-q^2v^2_Fr^2} 
- \frac{2}{d}\frac{\omega^2_p}{4\pi}~k~\left(1\mp e^{-kd}\right)~\frac{1-p}{2}. \label{59}
\end{eqnarray}
This expression explicitly shows that Im$\left[G^{+}_B(k,q;\omega_s)\right]=~$Im$\left[G^{-}_B(k,q;\omega_s)\right]<0$, leading to $\gamma_0>0$ by virtue of Eq.~(\ref{53}). 

It follows that $\mbox{Re}\left[G^{+/-}(k,q;\omega_s)\right] = (2/d)(\omega^2_p/4\pi)~k~\left(1\mp e^{-kd}\right)~(1+p)/2.$
Substituting this in Eq.~(\ref{52}) and converting the sum therein into an integral for large $d$, we get $\omega^{+/-}_s/\omega_p = \sqrt{1-\left(1\pm e^{-kd}\right)(1+p)/4}.$ See that $\omega^{+/-}_s$ depends on surface properties via the parameter $p$. Only for $p=1$ would the conventional value, $\omega_p/\sqrt{2}$, be recovered. For $p=0$ and at large $kd$, $\omega_s = \left(\sqrt{3}/2\right)\omega_p$ is slightly larger than the former. It is notable that, $\omega^{+}_s$ remains finite even for $k=0$, in distinct contrast with the Drude model. The reason is simple: in Drude model no electric field could exist in the metal for symmetric modes at $k=0$, while in our theory, due to a spatial spread of charge density, the electric field does not vanish. The same conclusion applies to semi-infinite metals. 

To estimate $\gamma_0$ by Eq.~(\ref{53}), we take in Eq.~(\ref{59}) $\int^1_0dr~\frac{q^2v^2_Fr^3}{\bar{\omega}^2-q^2v^2_Fr^2} \approx \left(qv_F/2\bar{\omega}\right)^2$ for simplicity. Thus, $\mbox{Im}\left[G^{+/-}_B(k,q;\omega_s)\right]\approx -(2/d)(\omega^2_p/4\pi)(3/4)(v_F/\omega_s)~(1+p)~q^2,$ which is then plugged in Eq.~(\ref{53}) to produce $\gamma_0 \sim (3/2\pi) (\omega_p/\sqrt{2})/(1\mp e^{-kd}) \approx 0.35~\omega_p.$ In obtaining this expression, we have put $\sum_l \frac{\left(q^{+/-}_l\right)^2}{\left(q^{+/-}_l\right)^2+k^2} \approx q_cd/2\pi$ with $q_cv_F \sim \omega_p/\sqrt{2}$. Landau damping has been excluded here, as the approximation only takes the real part of $\Omega(k,q;\omega_s)$. 

\textit{Numerical solutions}. We can also accurately solve Eq.~(\ref{51}) numerically. The results are displayed in Fig.~\ref{figure:f1} (a), (b) and (c). A comparison with the approximate solution is not direct, because the approximate solution has excluded   while the numerical solution has automatically taken care of Landau damping. It is stressed that, the numerical solutions do not depend on the value of $q_c$, provided it is large enough -- in excess of $k_s$.  

\section{Energy Conversion with surfaces}
\label{sec:energy}
In this section, we show that the surface plays a critical role in the energy conversion of bounded systems. While it might be straightforward to handle this issue if the surface potential $\phi_s(\mathbf{x},t)$ is exactly known, it is less clear otherwise. Here we derive from Eq.~(\ref{1}) a generic equation that governs the evolution of the electrostatic potential energy, denoted by $$E_p(t) = (1/2)\int d^3\mathbf{x}~\rho(\mathbf{x},t)\phi(\mathbf{x},t)$$ of the system, dispensing with the need to know $\phi_s(\mathbf{x},t)$. We then use it to furnish another proof of Eq.~(\ref{54}). For this purpose, we multiply Eq.~(\ref{1}) by $\phi(\mathbf{x},t)$ and integrate it over space to obtain
\begin{equation}
\left(\partial_t+\frac{2}{\tau}\right)E_p(t) = - ~P^{(1)}(t) - P^{(2)}(t),\label{3}
\end{equation}
where $P^{(1)}(t) = \int d^3\mathbf{x}~\mathbf{J}(\mathbf{x},t)\cdot\mathbf{E}(\mathbf{x},t)$ is no more than the work done by the electric field on the electrons per unit time and 
\begin{equation}
P^{(2)}(t) =  \frac{1}{2} ~\int d^3\mathbf{x}~\mathcal{J}(\mathbf{x},t)E_z(\mathbf{x},t), \quad \mathcal{J}(\mathbf{x},t) = J_z(\mathbf{x}_0,t)\Theta(z)-J_z(\mathbf{x}_d,t)\Theta(z-d). 
\end{equation}
It is evident that $P^{(2)}(t)$ signifies the work done by the surface on the electrons per unit time: electrons impinging toward the surface may lose their momentum. As far as we are concerned, this term and its consequences have hitherto not been discussed in existing work. We can translate Eq.~(\ref{3}) into the following, see Appendix \ref{sec:ec} for details, 
\begin{eqnarray}
&~&\gamma_0 ~ \int dz~\left(\mbox{Re}\left[\rho(z)\right]~\mbox{Re}\left[\phi(z)\right]+\mbox{Re}\rightarrow \mbox{Im}\right) \nonumber\\
&~&\quad \quad \quad \quad \quad \quad \quad \quad =-\frac{1}{2}\left\{\int dz~\left(\mbox{Re}\left[\mathbf{J}(z)\right]\cdot\mbox{Re}\left[\mathbf{E}(z)\right]+\mbox{Re}\rightarrow \mbox{Im}\right) +\frac{1}{2}\int dz~\left(\mbox{Re}\left[\mathcal{J}(z)\right]~\mbox{Re}\left[E_z(z)\right]+\mbox{Re}\rightarrow \mbox{Im}\right) \right\}, \label{8}
\end{eqnarray}
where the integral is extended over the metal. If the phase of $\rho(z)$ is global, i.e. independent of $z$, Eq.~(\ref{8}) holds valid even without the abbreviated term. 

Now we show how Eq.~(\ref{54}) can also be reached from Eq.~(\ref{8}). Neglecting Landau damping, by Eq.~(\ref{ch}) we can show that $\rho(z)$ has a global phase. We can then ignore in this equation the terms abbreviated as Re $\rightarrow$ Im without affecting the results. To the zeroth order in $\gamma_0$, it is obvious that Re$\left[\mathbf{J}_D(z)\right]\cdot\mbox{Re}\left[\mathbf{E}(z)\right] = 0$ and $\mbox{Re}\left[\mathcal{J}_{D}(z)\right]\mbox{Re}\left[E_z(z)\right]=0$, i.e. diffusive currents do not bear net work from the electric field. As for the surface-ballistic currents, note that $\mathbf{J}_B(\mathbf{v},z)$ contains the rapidly oscillating factor $e^{i\tilde{\omega}z/v_z}$, which suppresses the term $\int dz ~\mbox{Re}\left[\mathbf{J}_B(z)\right]\cdot\mbox{Re}\left[\mathbf{E}(z)\right]$ by the factor $kv_F/\omega_p$, echoing the fact that $\mathcal{M}^{+/-}$ can be neglected in Eq.~(\ref{47}). As such, we have
\begin{eqnarray}
\gamma_0 \approx -\frac{1}{2} \frac{\frac{1}{2} \int dz ~\mbox{Re}\left[\mathcal{J}_{B}(z)\right]~\mbox{Re}\left[E_z(z)\right]}{\int dz~ \mbox{Re}\left[\rho(z)\right]\mbox{Re}\left[\phi(z)\right]}
=-\frac{1}{4}~\frac{\mbox{Re}\left[J_{B,z}(0)\right]\mbox{Re}\left[\phi(0)\right]-\mbox{Re}\left[J_{B,z}(d)\right]\mbox{Re}\left[\phi(d)\right]}{\int dz ~\mbox{Re}\left[\rho(z)\right]\mbox{Re}\left[\phi(z)\right]} = -\frac{1}{2}\frac{\mbox{Re}\left[J_{z}(0)\right]\mbox{Re}\left[\phi(0)\right]}{\int dz~\mbox{Re}\left[\rho(z)\right]\mbox{Re}\left[\phi(z)\right]}, \nonumber
\end{eqnarray}
where in the last equality, we have used the fact that, for either symmetric or anti-symmetric modes $J_z(d)\phi(d) + J_z(0)\phi(0) = 0$, and that Re$\left[J_{z}(0)\right] = \mbox{Re}\left[J_{B,z}(0)\right]$. 
To evaluate the denominator, we utilize the equation of motion in real space. It can be easily obtained from Eq.~(\ref{1b}) with $\nabla\cdot\mathbf{J}(z) \approx \nabla\cdot\mathbf{J}_D(z) \approx (i/\bar{\omega})\omega^2_p\rho(z)$. We find 
\begin{equation}
\rho(z) \approx \frac{1}{i}\frac{\bar{\omega}}{\omega^2_p-\bar{\omega}^2}\left[J_{z}(d)\delta(z-d)-J_z(0)\delta(z)\right], \quad  \Longrightarrow \quad 
\mbox{Re}[\rho(z)] \approx \frac{\omega_s}{\omega^2_p-\omega^2_s} \left( \mbox{Im}\left[J_{z}(d)\right] \delta(z-d) - \mbox{Im}\left[J_{z}(0)\right]\delta(z)\right), 
\end{equation}
As a result, $\int dz~\mbox{Re}\left[\rho(z)\right]\mbox{Re}\left[\phi(z)\right] \approx -  \frac{\omega_s}{\omega^2_p-\omega^2_s}~\mbox{Im}\left[J_{z}(0)\right]\mbox{Re}\left[\phi(0)\right].$ By substitution, we immediately recover Eq.~(\ref{54}). 
\end{widetext}

\section{discussions and conclusions}
\label{sec:discussions}
Thus, on the basis of Boltzmann's equation, we have established a rigorous theory for SPWs in metal films with arbitrary electronic collision rate $1/\tau$. As a key consequence of the theory, we find that there exists a self-amplification channel for SPWs, which would cause the latter to spontaneously amplify at a rate $\gamma_0$ if not for electronic collisions. Surprisingly, the value of $\gamma_0$ turns out to be independent of $\tau$. The presence of this channel is guaranteed by the causality principle. Whether the system could actually amplify or not depends on the competition between $\gamma_0$ and $1/\tau$. If $\gamma_0>1/\tau$, SPWs will amplify and the system will become unstable. In our theory, the non-equilibrium deviation $g(\mathbf{v},z)$ refers to the Fermi-Dirac distribution $f_0(\varepsilon)$; as such, the instability is one of the Fermi sea. Needless to say, the instability will be terminated once the system deviates far enough from the Fermi sea and settles in a stable state. Clarifying the nature of the destination state is a subject of crucial importance for future study. 

One central feature of our theory is the classification of current densities into a diffusive component $\mathbf{J}_D(z)$ and a surface-ballistic component $\mathbf{J}_B(z)$. This classification is not based on the value of $\tau$ but according to whether the component obeys the (generalized) Ohm's law or not. Apart from this, these components are also discriminated in other ways. Firstly, they are controlled by different length scales. As it largely follows the local electric field $\mathbf{E}(z)$, the characteristic length associated with $\mathbf{J}_D(z)$ is $k^{-1}$. On the other hand, the length for $\mathbf{J}_B(z)$ is $v_F/\gamma_0$, because of simple $z$-dependence. Secondly, they are oriented disparately. $\mathbf{J}_D(z)$ is largely oriented normal to $\mathbf{E}(z)$ locally whereas $\mathbf{J}_B(z)$ normal to the surfaces -- especially for $p$ close to unity. Considering energy conversion, this explains why $\mathbf{J}_D(z)$ does not destabilize the Femi sea but $\mathbf{J}_B(z)$ does. Thirdly, $\mathbf{J}_D(z)$ is a bulk property and exists regardless of the surface; On the contrary, $\mathbf{J}_B(z)$ reflects true surface effects and it would disappear without surfaces. 

Although our theory applies at finite temperature, our calculation of $\bar{\omega}$ is done only at zero temperature, i.e. we have taken $f_0(\varepsilon)$ to be a step distribution. Clarifying the temperature dependence of Im$(\bar{\omega})$ is important for experimental studies of the present theory, because the net amplification/damping rate $\gamma = \gamma_0-1/\tau$ can be directly measured. Arguably, $\gamma_0$ could bear a different temperature dependence than $1/\tau$. In sufficiently pure samples, in which the residual resistivity is small enough, there might exist a critical temperature $T^*$, above which $\gamma<0$ while below it $\gamma>0$. In other words, $T^*$ marks the transition of the system from the Fermi sea to a more stable state. 

Another problem that needs to be addressed in the future for experimental studies is concerned with the effects of inter-band transitions. In the most experimented materials, such as silver and gold, these transitions are known to have dramatic effects. They not only open a loss channel due to inter-band absorption, but also significantly shift the SPW frequency. Including them in our formalism consists of a simple generalization: in addition to $\mathbf{J}_D(z)$ and $\mathbf{J}_B(z)$, the total current density $\mathbf{J}(z)$ must now also have a component $\mathbf{J}_{int}(z)$ accounting for inter-band transitions. The equation of motion is obtained by substituting $\mathbf{J}(z)$ in Eq.~(\ref{1b}). One may write $J_{int,\mu}(z) = \sum_{\nu}\int dz'~\sigma_{\mu\nu}(z,z';\omega)E_{\nu}(z')$, where $\mu, \nu = x,y,z$ and the inter-band conductivity $\sigma_{\mu\nu}$ can in principle be calculated using Greenwood-Kubo formula. In practice, calculating $\sigma_{\mu\nu}$ could be a formidable task even for the imaginably simplest surfaces. Nevertheless, one may argue that $\mathbf{J}_{int}(z)$ primarily affects the properties of bulk waves, namely, $\Omega(k,q;\bar{\omega})$. The causality principle should still protect the amplification channel, though the value of $\bar{\omega}$ may depend on $\tau$. A systematic analysis will be presented elsewhere. 

To conclude, we have presented a theory for SPWs in metal films taking into account the unique interplay between ballistic electronic motions and boundary effects, from which it emerges a universal self-amplification channel for these waves. It is expected that the study will bear far-reaching practical and fundamental consequences, which are to be explored in the future. We hope that the work could stimulate more effort on this subject.  

\section*{acknowledgement}
The author is grateful to K. Wakabayashi for some help with numerical computation. He also thanks E. Mariani for enormous support. 

\begin{widetext}
\appendix
\section{More about Eqs.~(\ref{3}) - (\ref{8})}
\label{sec:ec}
The not-so-obvious step in proving Eq.~(\ref{8}) is to show that 
\begin{equation}
\partial_t E_p(t) = \int d^3\mathbf{x}~\phi(\mathbf{x},t)\partial_t\rho(\mathbf{x},t). 
\end{equation}
For this purpose, we write $\phi(\mathbf{x},t) = \mbox{Re}\left[e^{-i\omega t}\phi(\mathbf{x})\right]$ and $\rho(\mathbf{x},t) = \mbox{Re}\left[e^{-i\omega t}\rho(\mathbf{x})\right]$, where $\phi(\mathbf{x}) = e^{ikx}\phi(z)$ and $\rho(\mathbf{x}) = e^{ikx}\rho(z)$. Moreover, we put $\omega = \omega_s + i\gamma$, $\phi(\mathbf{x}) = \phi'(\mathbf{x}) + i\phi''(\mathbf{x})$ and similarly for other complex quantities. By substitution, we find 
\begin{eqnarray}
\partial_tE_p(t) ~ - \int d^3\mathbf{x}~\phi(\mathbf{x},t)\partial_t\rho(\mathbf{x},t) = \frac{e^{2\gamma t}}{2}\int d^3\mathbf{x}~\left[\phi''(\mathbf{x})\rho'(\mathbf{x})-\phi'(\mathbf{x})\rho''(\mathbf{x})\right]. 
\end{eqnarray}
However, $\int d^3\mathbf{x} ~ \phi'(\mathbf{x})\rho"(\mathbf{x}) = \int d^3\mathbf{x}~\phi''(\mathbf{x})\rho'(\mathbf{x}) = 0$. Actually, we have
\begin{eqnarray}
&~&\int d^3\mathbf{x} ~ \phi'(\mathbf{x})\rho"(\mathbf{x}) 
=\int dz \int d^2\mathbf{r} \left(\phi'(z)\rho"(z)\cos^2kx-\phi"(z)\rho'(z)\sin^2kx\right) \nonumber\\
&\propto& \int dz \left(\phi'(z)\rho"(z)-\phi"(z)\rho'(z)\right)
\propto \int dz \int dz' \left[\rho"(z)e^{-k\abs{z-z'}}\rho'(z') - \rho'(z)e^{-k\abs{z-z'}}\rho"(z')\right]=0, 
\end{eqnarray}
thus completing the proof. 

Let us suppose $\rho(z)$ has a global phase, i.e. $\rho(z) = c~n(z)$, where $c$ is a complex constant and $n(z)$ is real-valued. One can show that Eq.~(\ref{8}) can be turned into an equation that involves only $n(z)$, wherein $c$ plays no role. In other words, Eq.~(\ref{8}) can be evaluated by simply pretending $c$ (and $\rho(z)$) to be real. The proof is evident considering the linear relations between $\rho(z)$ and $\mathbf{J}(z)$ and that between $\rho(z)$ and $\phi(z)$ as well as that between $\rho(z)$ and $\mathbf{E}(z)$. 

\section{Electronic distribution functions}
\label{sec:edf}
The general solution to Eq.~(\ref{11}) is given by 
\begin{equation}
g(\mathbf{v},z) = e^{i\frac{\tilde{\omega}z}{v_z}}\left(C(\mathbf{v})-\frac{e\partial_{\mathbf{v}}f_0}{mv_z}\cdot \int^z_0~dz'~e^{-i\frac{\tilde{\omega}z'}{v_z}}\mathbf{E}(z')\right), \label{b1}
\end{equation}
where $C(\mathbf{v})$ is an arbitrary integration constant to be determined by boundary conditions. Let $p_1$ and $p_2$ be the \textit{Fuchs} parameters for the (uniform) surfaces at $z=0$ and $z=d$, respectively. The boundary condition at $z=0$ is taken that $g\left((v_x,v_y,v_z>0),z=0\right)=p_1~g\left((v_x,v_y,-v_z),z=0\right)$ while that at $z=d$ assumes $g\left((v_x,v_y,v_z<0),z=0\right)=p_2~g\left((v_x,v_y,-v_z),z=0\right)$, both evaluated at $E_z(z) = 0$. After some algebra, one finds
\begin{eqnarray}
g(\mathbf{v},z) = e^{i\frac{\tilde{\omega}z}{v_z}}
\begin{cases}
\frac{1}{p_1p_2-e^{-\frac{2d\tilde{\omega}}{v_z}}}\int^d_0dz'~\frac{e\mathbf{E}(z')\cdot\partial_{\mathbf{v}}f_0}{mv_z}\left(e^{-i\frac{(2d+z')\tilde{\omega}}{v_z}}+p_1e^{-i\frac{(2d-z')\tilde{\omega}}{v_z}}\right) + \int^d_z dz'~\frac{e\mathbf{E}(z')\cdot\partial_{\mathbf{v}}f_0}{mv_z} e^{-i\frac{\tilde{\omega}z'}{v_z}}, & \mbox{for}~v_z\geq 0,\\
\frac{-1}{p_1p_2-e^{\frac{2d\tilde{\omega}}{v_z}}}\int^d_0dz'~\frac{e\mathbf{E}(z')\cdot\partial_{\mathbf{v}}f_0}{mv_z}\left(e^{i\frac{(2d-z')\tilde{\omega}}{v_z}}+p_1e^{i\frac{\tilde{\omega}z'}{v_z}}\right) - \int^z_0 dz'~\frac{e\mathbf{E}(z')\cdot\partial_{\mathbf{v}}f_0}{mv_z} e^{-i\frac{\tilde{\omega}z'}{v_z}}, & \mbox{for}~v_z< 0.
\end{cases}
\end{eqnarray}
The electronic distribution functions presented in the main text in Sec.~\ref{sec:methods} are obtained by approximating $\left(1-p_1p_2e^{i\frac{2d\tilde{\omega}}{v_z}}\right)^{-1}\approx 1$ for $v_z\geq 0$ and $\left(1-p_1p_2e^{-i\frac{2d\tilde{\omega}}{v_z}}\right)^{-1}\approx 1$ for $v_z< 0$ in this equation. 

\textit{Causality principle}. It should be pointed out that, in applying the boundary conditions, we have implicitly assumed Im$(\tilde{\omega})\geq 0$; otherwise, we would find unphysical solutions that violate the principle of causality, which states that the number of out-going electrons is determined by the number of in-coming electrons, not otherwise. It is easy to show that, had we assumed Im$(\tilde{\omega})<0$, we would have found the opposite: the number of reflected electrons would be fixed while the number of incident electrons would go to infinity as $p_{1/2}\rightarrow 0$. 

\section{The matrix $\mathcal{M}^{+/-}$}
\label{sec:M}
In the first place, we show that $\mathcal{M}^{+/-}/\omega^2_p\propto i kv_F/\bar{\omega} + ...$, where the ellipsis stands for higher order terms in $kv_F/\bar{\omega}$. We take the symmetric modes for illustration, as the reasoning can be replicated for the anti-symmetric modes as well. Writing $\int d^3\mathbf{v} ~\Theta(v_z) = \int^{2\pi}_0d\varphi \int^{\pi/2}_0d\theta \sin\theta \int^{\infty}_0 dv^2 (v/2)$ and integrating over $v$, we find 
\begin{eqnarray}
\frac{\mathcal{M}^{+}_{l,l'}}{\omega^2_p} &=& i\left(1-\frac{\delta_{l,0}}{2}\right) \frac{3}{2\pi kd}\int^{2\pi}_0d\varphi \int^{\pi/2}_0d\theta \sin\theta \frac{\bar{\omega}^2\cos\theta-\bar{\omega}kv_F\sin\theta\cos\theta\cos\varphi}{(\bar{\omega}-kv_F\sin\theta\cos\varphi)^2-\left(q^+_{l}\right)^2v^2_F\cos^2\theta}\frac{k\bar{\omega}/v_F}{k^2+\left(q^+_{l'}\right)^2}\nonumber\\&~&\times\left(e^{i\left(\frac{\bar{\omega}}{v_F\cos\theta}-k\tan\theta\cos\varphi\right)d}-1\right) L(v_F\sin\theta\cos\varphi,v_F\cos\theta,k,q^+_{l'},\bar{\omega},+1). \label{c1}
\end{eqnarray}
To the lowest order in $kv_F/\bar{\omega}$, we only need to retain $L^{(0)}$ in the expansion $L^{sym} = \sum^{\infty}_{m=0}L^{(m)}\left(kv_F/\bar{\omega}\right)^m$. Thus, 
\begin{equation}
L(v_F\sin\theta\cos\varphi,v_F\cos\theta,k,q,\bar{\omega},+1) \approx 2~\frac{q^2v^2_F\cos^2\theta}{\bar{\omega}^2-q^2v^2_F\cos^2\theta}~\left(1+p-p~e^{i\left(\frac{\bar{\omega}}{v_F\cos\theta}-k\tan\theta\cos\varphi\right)d}\right).
\end{equation}
Substituting this back in (\ref{c1}) and approximating
\begin{equation}
\frac{\bar{\omega}^2\cos\theta-\bar{\omega}kv_F\sin\theta\cos\theta\cos\varphi}{(\bar{\omega}-kv_F\sin\theta\cos\varphi)^2-q^2v^2_F\cos^2\theta} \approx \frac{\bar{\omega}^2\cos\theta}{\bar{\omega}^2-q^2v^2_F\cos^2\theta}, \quad \frac{q^2}{q^2+k^2} \approx 1,
\end{equation}
we arrive at 
\begin{equation}
\frac{\mathcal{M}^{+}_{l,l'}}{\omega^2_p} = i\left(1-\frac{\delta_{l,0}}{2}\right) \frac{3}{\pi kd}\left(\frac{kv_F}{\bar{\omega}}\right)\int^{2\pi}_0d\varphi \int^{\pi/2}_0d\theta \sin\theta~ \frac{\cos^3\theta}{1-(q^+_{l}v_F/\bar{\omega})^2\cos^2\theta}  \frac{\mathcal{L}(\theta,\varphi)}{1-(q^+_{l'}v_F/\bar{\omega})^2\cos^2\theta},
\end{equation}
where
\begin{equation}
\mathcal{L}(\theta,\varphi) = \left(e^{i\left(\frac{\bar{\omega}}{v_F\cos\theta}-k\tan\theta\cos\varphi\right)d}-1\right) \times \left(1+p-p~e^{i\left(\frac{\bar{\omega}}{v_F\cos\theta}-k\tan\theta\cos\varphi\right)d}\right).
\end{equation}
Clearly, we have $\mathcal{M}^{+}/\omega^2_p \sim kv_F/\bar{\omega}$, as stated.

We may proceed further If we take
\begin{equation}
\frac{3}{\pi} \int^{2\pi}_0d\varphi \int^{\pi/2}_0d\theta \sin\theta \frac{\cos^3\theta}{1-(q^+_{l}v_F/\bar{\omega})^2\cos^2\theta}  \frac{\mathcal{L}(\theta,\varphi)}{1-(q^+_{l'}v_F/\bar{\omega})^2\cos^2\theta} \approx \frac{3}{\pi} \int^{2\pi}_0d\varphi \int^{\pi/2}_0d\theta \sin\theta ~\cos^3\theta ~ \mathcal{L}(\theta,\varphi) \sim -1,
\end{equation}
from which it follows that $\mathcal{M}^{+}_{l,l'} \approx M_0 = -i\omega^2_p(1/kd)(kv_F/\bar{\omega})$, which is a constant. Therefore, $\mathcal{M}^{+}_{l,l'} \approx M_0\mathbb{Z}_{l,l'}$, where $\mathbb{Z}_{l,l'}=1$ constitutes a unity matrix. We write, with $\mathcal{W}^+_{l,l'} = \delta_{l,l'}\Omega(k,q^+_l;\bar{\omega})$, 
\begin{equation}
\left[\left(\mathcal{W}^{+}\right)^2-\bar{\omega}^2\mathbb{I}+\mathcal{M}^{+}\right]^{-1} = U^{-1}\left[\left(\tilde{\mathcal{W}}^{+}\right)^2-\bar{\omega}^2\mathbb{I}+\tilde{\mathcal{M}}^{+}\right]^{-1}U,
\end{equation}
where $U$ is a similarity transformation that brings $\mathcal{M}^{+}$ and hence $\mathbb{Z}$ to a diagonal form. We have used a tilde to indicate the transformed matrices, e.g. we write $\tilde{\mathbb{Z}} = U\mathbb{Z}U^{-1}$. See that $\tilde{\mathbb{Z}}$ has only one non-vanishing element, whose value amounts to the dimension $N_c$ of the matrix. Let it be the $l_0$-th element. Then $\tilde{\mathbb{Z}}_{l,l'} = N_c\delta_{l,l_0}\delta_{l',l_0}$. Obviously, $N_c = q_cd/2\pi \sim (\bar{\omega}/kv_F)(kd/2\pi) $. As such, $M_0\sim 1/N_c$ and $\tilde{\mathcal{M}}^{+}_{l,l'}\sim -i(\omega^2_p/2\pi) \delta_{l,l_0}\delta_{l',l_0}$. Introducing $\tilde{\mathcal{G}}^{+}=\mathcal{G}^{+}U^{-1}$ and $\tilde{\mathbb{E}}_{+} = U\mathbb{E}_{+}$, we can rewrite the equation of motion for the symmetric modes as 
\begin{eqnarray}
1 = \tilde{\mathcal{G}}^{+} \left[\left(\tilde{\mathcal{W}}^{+}\right)^2-\bar{\omega}^2\mathbb{I}+\tilde{\mathcal{M}}^{+}\right]^{-1} \tilde{\mathbb{E}}_{+}.
\end{eqnarray}
Taking $\mathcal{W}^+\approx \omega_p\mathbb{I}$ and hence $\tilde{\mathcal{W}}^{+}\approx \omega_p\mathbb{I}$, this equation becomes
\begin{eqnarray}
1 = \sum_l \tilde{\mathcal{G}}^{+}_l \frac{1}{\omega^2_p-\bar{\omega}^2+\tilde{\mathcal{M}}_{l,l}^{+}}\tilde{\mathbb{E}}_{+,l} =  \sum_l \mathcal{G}^{+}_l \frac{1}{\omega^2_p-\bar{\omega}^2}\mathbb{E}_{+,l} + \left[\tilde{\mathcal{G}}^{+}_{l_0} \frac{1}{\omega^2_p(1+i/2\pi)-\bar{\omega}^2}\tilde{\mathbb{E}}_{+,l_0} - \mathcal{G}^{+}_{l_0}\frac{1}{\omega^2_p-\bar{\omega}^2}\mathbb{E}_{+,l_0}\right] \approx \sum_l \mathcal{G}^{+}_l \frac{1}{\omega^2_p-\bar{\omega}^2}\mathbb{E}_{+,l}. \nonumber
\end{eqnarray}
The term in the square bracket makes only a contribution of the order of $\sim 1/N_c$ and can be neglected for large $N_c$. 
\end{widetext}


\begin{thebibliography}{9}

\bibitem{ritchie1957}
R. H. Ritchie, Phys. Rev. \textbf{106}, 874 (1957).

\bibitem{ferrell1958}
R. A. Ferrell, Phys. Rev. \textbf{111}, 1214 (1958).

\bibitem{raether1988}
H. Raether, \textit{Surface plasmons on smooth and rough surfaces and on gratings} (Springer Berlin Heidelberg,1988).

\bibitem{pitarke2007}
J. M. Pitarke, V. M. Silkin, E. V. Chulkov and P. M. Echenique, Rep. Prog. Phys. \textbf{70}, 1 (2007).

\bibitem{feibelman1982}
P. J. Feibelman, Prog. Surf. Science. \textbf{12}, 287 (1982).

\bibitem{pendry1975}
P. M. Echenique and J. B. Pendry, J. Phys. C: Solid State Phys. \textbf{8}, 2936 (1975).

\bibitem{int}
Interview with J. Krenn, Nat. Photonics \textbf{6}, 714 (2012).

\bibitem{benno1988}
B. Rothenh\"{a}usler and K. Wolfgang, Nature \textbf{332}, 615 (1988).

\bibitem{hoa2007}
X. D. Hoa, A. G. Kirk and M. Tabrizian, Biosensors and Bioelectronics \textbf{23}, 151 (2007).

\bibitem{zayats2005}
A. V. Zayats, I. S. Igor and A. A. Maradudin, Phys. Rep. \textbf{408}, 131 (2005).

\bibitem{maier2007}
S. A. Maier \textit{Plasmonics: fundamentals and applications} (Springer Science \& Business Media, 2007).

\bibitem{ozbay2006}
E. Ozbay, Science \textbf{311}: 189 (2006).

\bibitem{mark2007}
M. L. Brongersma and P. G. Kik, \textit{Surface plasmon nanophotonics} (Springer, 2007).

\bibitem{barnes2003}
W. L. Barnes, A. Dereux, and T. W. Ebbesen, Nature \textbf{424}, 824 (2003); W. L. Barnes, J. of Optics A \textbf{8}, S87 (2006).

\bibitem{ebbesen2008}
T. W. Ebbesen, C. Genet and S. I. Bozhevolnyi, Physics Today \textbf{61}, 44 (2008).

\bibitem{tame2013}
M. S. Tame, K. R. McEnergy, S. K. Ozdemir, J. Lee, S. A. Maier and M. S. Kim, Nat. Phys. \textbf{9}, 329 (2013).

\bibitem{nie1997}
S. Nie and Steven R. Emory, Science \textbf{275}, 1102 (1997).

\bibitem{sarid2010}
D. Sarid and W. Cgallener, \textit{Modern introduction to surface plasmons: theory, mathematic modeling and applications} (Cambridge University Press, Cambridge, UK, 2010).

\bibitem{harris1971}
J. Harris, Phys. Rev. B \textbf{4}, 1022 (1971).

\bibitem{fetter1986}
A. L. Fetter, Ann. Physics \textbf{81}, 367 (1973); Phys. Rev. B \textbf{33}, 3717 (1986).

\bibitem{pendry2013}
Y. Luo, A. I. Fernandez-Dominguez, A. Wiener, S. A. Maier and J. B. Pendry, Phys. Rev. Lett. \textbf{111}, 093901 (2013).

\bibitem{pendry2014}
Y. Luo, R. Zhao and J. B. Pendry, Proc. Natl. Acad. Sci. U.S.A. \textbf{111}, 18422 (2014). 

\bibitem{pendry2015}
J. B. Pendry, Y. Luo and R. Zhao, Science \textbf{348}, 521 (2015)

\bibitem{schnitzer2016}O. Schnitzer, V. Giannini, S. A. Maier and R. V. Craster, Proc. R. Soc. A 472, 20160258 (2016). 

\bibitem{ziman1960}
J. M. Ziman, \textit{Electrons and Phonons: the theory of transport phenomena in solids} (Oxford University Press, 2001).

\bibitem{pines}
D. Pines, \textit{Elementary excitations in solids} (W. A. Benjamin, New York, 1963)

\bibitem{abrikosov}
A. A. Abrikosov, \textit{Fundamentals of the theory of metals} (Elsiver Science Publishers B. V., North-Holland, 1988)

\bibitem{deng2017a}
H.-Y. Deng, K. Wakabayashi and C.-H. Lam, Phys. Rev. B \textbf{95}, 045428 (2017).

\bibitem{deng2017b}
H.-Y. Deng, arXiv:1606.06239 (2016). 

\bibitem{deng2015}
H.-Y. Deng, and K. Wakabayashi, Phy. Rev. B \textbf{92}, 045434 (2015).

\bibitem{bergman2003}
D. J. Bergman and M. I. Stockman, Phys. Rev. Lett. \textbf{90}, 027402 (2003).

\bibitem{seidel2005}
J. Seidel, S. Frafstron and L. Eng, Phys. Rev. Lett.\textbf{94}, 177401 (2005).

\bibitem{leon2008}
I. De Leon and P. Berini, Phys. Rev. B \textbf{78}, 161401(R) (2008).

\bibitem{leon2010}
I. De Leon and P. Berini, Nat. Photonics \textbf{4}, 382 (2010).

\bibitem{yu2011}
D. Y. Fedyanin and A. Y. Arsenin, Opt. Express \textbf{19}, 12524 (2011).

\bibitem{pierre2012}
P. Berini and I. De Leon, Nat. Photonics \textbf{6},16 (2012).

\bibitem{fedyanin2012}
D. Y. Fedyanin, A. V. Arsenin and A. V. Zayats, Nano Letters \textbf{12}, 2459 (2012).

\bibitem{cohen2013}
S. K\'{e}na-Cohen, P. N. Stavrinou, D. D. C. Bradley and S. A. Maier, Nano Letters \textbf{13}, 1323 (2013).

\bibitem{brown1960}
R. W. Brown, P. Wessel and E. P. Trounson, Phys. Rev. Lett. \textbf{5}, 472 (1960).

\bibitem{arakawa1964}
E. T. Arakawa, R. J. Herickhoff and R. D. Birkhoff, Phys. Rev. Lett. \textbf{12}, 319 (1964).

\bibitem{donohue1986}
J. F. Donohue and E. Y. Wang, J. Appl. Phys. \textbf{59}, 3137 (1986); \textbf{62}, 1313 (1987).


\end{thebibliography}
\end{document}